\documentclass{aa}  

\usepackage{graphicx}

\usepackage[colorlinks, linkcolor=blue, anchorcolor=blue, citecolor=blue]{hyperref}

\usepackage{txfonts}
\usepackage{natbib}
\bibpunct{(}{)}{;}{a}{}{,} 

\begin{document}

   \title{Observable tests for the light-sail scenario of interstellar objects}

\author{Wen-Han Zhou
          \inst{1,2}
          \and
          Shang-Fei Liu
          \inst{3,4}
          \and
          Yun Zhang
          \inst{5}
          \and
          Douglas N.C. Lin
          \inst{6,7}
          }

   \institute{Universit\'e C\^ote d'Azur, Observatoire de la C\^ote d'Azur, CNRS, Laboratoire Lagrange, Nice 06304, France\\
              \email{wenhan.zhou@oca.eu}
        \and
        Origin Space Co. Ltd., China 
        \and
        School of Physics and Astronomy, Sun Yat-sen University, 2 Daxue Road, Zhuhai 519082, Guangdong Province, China\\
               \email{liushangfei@mail.sysu.edu.cn}
        \and
        CSST Science Center for the Guangdong-Hong Kong-Macau Greater Bay Area, Sun Yat-sen University, 2 Daxue Road, Zhuhai 519082, Guangdong Province, China
        \and
        Department of Aerospace Engineering, University of Maryland, College Park, MD, 20742, USA
        \and
        Department of Astronomy and Astrophysics, University of California, Santa Cruz, USA
        \and
        Institute for Advanced Studies, Tsinghua University, Beijing, 100086, China
             }

  \abstract
   {Enigmatic dynamical and spectral properties of the first interstellar 
object (ISO), 1I/2017 U1 (`Oumuamua), led to many hypotheses, including a suggestion that it may be an ``artificial'' spacecraft with a thin radiation-pressure-driven light sail. Since similar discoveries by forthcoming instruments, such as the Vera Rubin telescope and the Chinese Space Station Telescope (CSST), are anticipated, a critical identification of key observable tests is warranted for the quantitative distinctions between various scenarios.}
   {We scrutinize the light-sail scenario by making comparisons between physical models and observational constraints. These analyses can be generalized for future surveys of `Oumuamua-like objects.}
  {The light sail goes through a drift in interstellar space due to the magnetic field and gas atoms, which poses challenges to the navigation system. When the light sail enters the inner Solar System, the sideways radiation pressure leads to a considerable non-radial displacement. The immensely high dimensional ratio and the tumbling motion could cause a light curve with an extremely large amplitude and could even make the light sail invisible from time to time. These observational features allow us to examine the light-sail scenario of interstellar objects.}
   {The drift of the freely rotating light sail in the interstellar medium is $\sim 100\,$au even if the travel distance is only 1 pc. The probability of the expected
brightness modulation of the light sail matching with `Oumuamua's observed variation amplitude ($\sim$ 2.5 -- 3) is < 1.5\%.  In addition, the probability of the tumbling light sail being  visible (brighter than V=27) in all 55 observations spread over two months after discovery is 0.4\%. Radiation pressure could cause a larger displacement normal to the orbital plane for a light sail than that for `Oumuamua. Also, the ratio of
antisolar to sideways acceleration of `Oumuamua deviates from that of the light sail by  $\sim $1.5 $\sigma$. }
   {We suggest that `Oumuamua is unlikely to be a light sail. The dynamics of an intruding light sail, if it exists, has distinct observational signatures, which can be quantitatively identified and analyzed with our methods in future surveys.}


   \maketitle
%

\section{Introduction}
\label{sec:1}

The unprecedented trajectory and unique spectral properties \citep{Meech17} exhibited during a brief visit of the first interstellar object (ISO), 1I/2017 U1 (`Oumuamua), gave rise to a protracted controversy regarding its origin \citep{Raymond18a, Raymond18b, Cuk18, Moro-Martin2019, Zhang20, Jackson21}.  According to one particularly bold assumption, it is an ``artificial'' spacecraft with a thin light sail driven by radiation pressure on an intentional mission to 
our Solar System \citep{BialyLoeb18, Loeb21}. 
Based on {the sensitivity limitation of the Panoramic Survey Telescope and Rapid Response System (Pan-STARRS)}, the interplanetary density of `Oumuamua-like ISOs is estimated 
to be $n_{\rm ISO} \lesssim 0.2$ au$^{-3}$ in the proximity of the Earth 
orbit \citep{Do18, Trilling17, Bannister19}.  
If their parent bodies randomly came across the Solar System, the 
inferred mass of ejected objects per star, $M_{\rm ej}$, would exceed a few $M_\oplus$ in the solar 
neighborhood \citep{Do18,Portegies18}.  Prior to its entry 
into the Sun's domain of gravitational influence, `Oumuamua's velocity relative 
to the local standard of 
rest (LSR) was $v_{\rm O} \sim 10$ km s$^{-1}$. At a heliocentric distance 
$r\sim 1.5$ au, its large-amplitude, irregularly varying light curve \citep{Meech17,Drahus18,Bolin17,Bannister17,Knight17,Jewitt17}
suggests that `Oumuamua is either a prolate or oblate 
object, freely tumbling with  {the period of its light curve increasing at a rate ${\dot P}_{\rm O} \sim 0.3$ hour/day from $P_{\rm O} \sim 8$ hours \citep{Flekkoy19}.} In the absence
of any detectable trace of outgassing, its orbital path has an observationally
inferred  {perihelion} distance of $p_{\rm O} \simeq 0.26$ au and departs from a 
parabolic Keplerian trajectory with a component of nongravitational 
acceleration predominantly in the radial direction away from the Sun \citep{Micheli18},  {although the validity of the fitting procedure has been questioned by \citet{Katz2019}}. 
Its surface albedo resembles that of asteroids rather than 
comets \citep{Trilling18}. Shortly after `Oumuamua was discovered, data taken with multiple telescopes
provided a high-cadence light curve ({from October 25, 2017, to October 27, 2017}) with an irregular brightness variation over an amplitude range $\sim 2.5-3$.  This high-cadence light curve has been used to deduce `Oumuamua's tumbling motion \citep{Meech17}. In addition, `Oumuamua was detected \citep{Micheli18} with magnitudes brighter than 27 in 
all 55 follow-up trackings (over 49 days starting  {on November 15, 2017}).

A rationale for the spacecraft hypothesis is that it circumvents the large
$M_{\rm ej}$ inferred for the parent bodies of randomly distributed ISOs.  It also attributes `Oumuamua's nongravitational acceleration to the 
Sun's radiation pressure to a light sail under the assumption that 
it has a thin oblate shape with a length $l_1 \sim 10^2$ m, width 
$l_2 \simeq l_1$, surface area $A$, mass $m_{\rm O}$, and surface 
density \citep{BialyLoeb18}
${m_{\rm O} / A} = \rho_{\rm O}  l_3 \approx 1 {\rm kg \ m} ^{-2}$. 
If `Oumuamua's density, $\rho_{\rm O} $, is comparable to that of refractory asteroids ($\simeq 2-3 
\times 10^3$ kg m$^{-3}$), its effective thickness, $l_3 (\sim 3-5 \times 10^{-4} {\rm m} < < l_1)$, would 
be comparable to that of an eggshell, two orders of magnitude smaller than that of the
International Space Station; although, a much thinner light sail  $(\sim 7 \times 10^{-6} {\rm m}$) 
has been deployed, without tumbling, on the spacecraft {Interplanetary Kite-craft Accelerated by Radiation Of the Sun (IKAROS)} \citep{Tsuda11}. 

However, the turbulent interstellar medium (ISM) poses challenges to the navigation of the light sail. Hydrogen atoms exert a drag force that is normal to the surface of the light sail, leading to a {sideways} acceleration of the tumbling light sail. Moreover, in the disorderly magnetic field, ${\vec B}$, of interstellar space, a Lorentz force in the direction $( {\vec B} \cdot {\vec v}_{\rm mag}
{\vec B} - \vert {\vec B} \vert^2 {\vec v}_{\rm mag})/\vert {\vec B} 
\vert ^2 \vert {\vec v}_{\rm mag} \vert$ is induced, which contains a drag and a sideways 
force component  with a magnitude $a_{\rm mag} \sim B^2 v_{\rm mag} / 
2\rho_{O} l_3 \mu_0 v_{\rm A}$, where $v_{\rm A}$ is the Alfv{\'e}n 
speed in the ISM. As `Oumuamua passes through randomly
oriented ${\vec B}$ in eddies of all sizes, $L_{\rm tur}$, the magnitude 
and direction of the magnetic drag change on a timescale $\tau_{\rm tur}
\sim L_{\rm tur}/v_{\rm O}$.  On timescales $t \geqslant \tau_{\rm tur}$, the value of `Oumuamua's path deviation from its initial target goes through a random walk.

Even if the light sail manages to arrive at the Solar System, the unique configuration (i.e., the extremely high dimensional ratio) of the light sail would give rise to recognizable observational features. The radiation pressure would cause a sideways acceleration due to the imperfect alignment of the surface normal vector with heliocentric position vector, which would result in a considerable non-radial deviation from Kepler motion. In addition, the tumbling of the extremely thin light sail would lead to a light curve with an  {immense} amplitude. The light sail can even be invisible when the sunlight is edge-on to the light sail. Stringent probability assessments for the spaceship scenario can be made using `Oumuamua's observed light curve and detected magnitude as well as a prefect recurrence rate.

 {Previous studies \citep[e.g.,][]{Katz2021,Curran2021} have questioned the light-sail scenario based on the unlikely coincidence between the advancement of extraterrestrial technology and the advent of the observational capability of Pan-STARRS, though a quantitative and comprehensive analysis of observation data has not yet been performed.} In this paper we revisit the light-sail scenario with its model parameters and
estimate the ram-pressure and magnetic drag and the torque on the light sail during `Oumuamua’s voyage through the turbulent ISM (Sect. \ref{sec:2}). 
In Sect. \ref{sec:3} we model `Oumuamua's dynamics as a tumbling light sail cruising the Solar System, and in particular we study the implications of its nongravitational acceleration 
due to the sideways radiation force and the observable features due to its tumbling motion and high dimensional ratio. We summarize our results and compare them with observations in Sect. \ref{sec:4}. 
Lastly, we discuss our results and propose observable tests for future surveys in Sect. \ref{sec:5}.


\section{Dynamics of a tumbling light sail in interstellar space}
\label{sec:2}

In the solar neighborhood, the ISM contains \citep{Draine10}: (1) a diffuse warm neutral 
medium (WNM) with 
number density $n_{\rm ISM} \sim 10^6$ m$^{-3}$ and sound speed $c_{\rm s} \sim 9$ km 
s$^{-1}$; (2) a cold neutral medium (CNM) with $n_{\rm ISM} 
\sim 10^8$ m$^{-3}$, $c_{\rm s} \sim 1.3$ km s$^{-1}$; and (3) giant 
molecular clouds (GMCs) with $n_{\rm ISM} \sim 10^{10}$--$10^{12}$ 
m$^{-3}$, $c_{\rm s} \sim 0.5$ km s$^{-1}$. The local
volume filling factor for the WNM, the CNM, and GMCs is 0.5, $5 \times 10^{-3}$, and 
$2 \times 10^{-3}$, respectively \citep{Draine10}, although `Oumuamua may have recently 
traveled through some GMCs \citep{Pfalzner20, Hsieh21}.  Due to the perturbation of GMCs, the velocity dispersion of  `Oumuamua and
nearby stars relative to the LSR $v_d (\tau_\ast) (\sim 10 
\sqrt{{\tau_\ast}/{1~\mathrm{Gyr}}}~\mathrm{km/s}$) increases as they advance in age, 
$\tau_\ast$. Their corresponding rate of kinetic energy gain, per unit mass, is 
${\dot e}_{\rm gain} = (1/2) d v_d^2/d\tau_\ast \sim 50~ \mathrm{km}^2
\mathrm{s}^{-2} \mathrm{Gyr}^{-1}$. 

\begin{figure*}[htbp]
    \begin{center}
    \includegraphics[width=\linewidth]{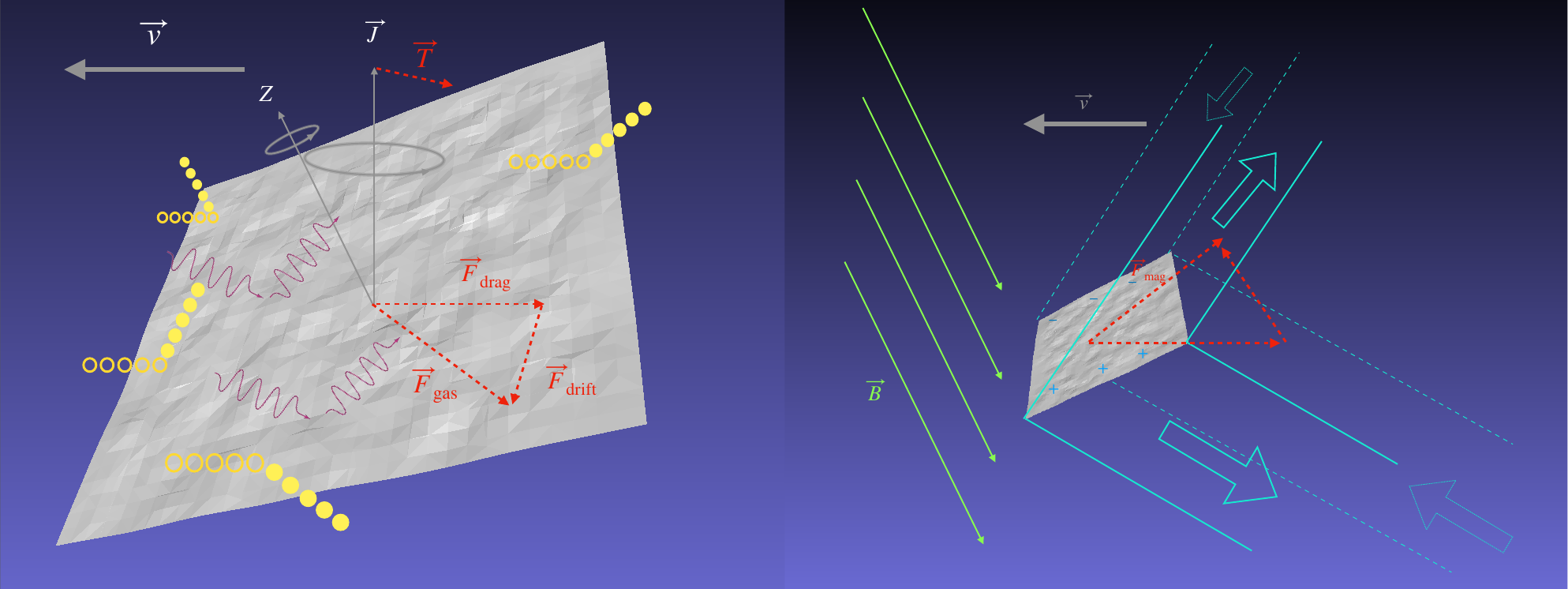}
    \caption{Schematic illustration of the hydrodynamic (left panel) and magnetic (right panel) drag on a moving light sail in a turbulent 
ISM with a magnetic field, $\vec B$ (green). The tumbling sail is represented by a thin gray sheet with 
a bumpy surface and a spin axis in the $Z$ direction that precesses around its angular momentum, 
$\vec J$, axis.  The bombardment of atoms (with open and solid yellow dots representing pre- and post-collision trajectories) and photons (wavy arrows) leads to a force, ${\vec F}_{\rm gas}$ (red), which 
can be decomposed into a drag, ${\vec F}_{\rm drag}$, in the direction of the sail's velocity, 
$\vec v$ (gray), and a sideways force, ${\vec F}_{\rm drift}$.  {The induced currents (blue arrows) conducted away by surrounding charged particles} lead to a Lorentz force, ${\vec  F}_{\rm mag}$, 
with components  {perpendicular} to both $\vec v$ and $\vec B$.
}
    \label{fig:solarsail_rotation}
    \end{center}
\end{figure*}

With a relative velocity $v$, atoms in the volume-filling WNM bombard the 
tumbling light sail. Their momentum impulse on its bumpy surface introduces
a residual torque and modulates its rotation frequency \citep{Zhou20} with a chaotic 
reorientation of its spin axis (Fig. \ref{fig:solarsail_rotation}) 
on a timescale $\tau_{\rm{rot}}$. The efficiency of
this effect can be extrapolated \citep{Mashchenko19} from ${\dot P}_{\rm O}$ 
due to a similar torque induced by the solar radiation on the light sail, 
though the required surface roughness is $\sim 10^2 l_3$ (i.e., much larger than its thickness). Readers are referred to Appendix 
\ref{sec:ismtorque} for our detailed analysis on the ISM's torque on the 
light sail's asymmetric surface and its application to `Oumuamua's spin.

Collectively, atoms in the streaming WNM exert a drag on the hypothetical tumbling light 
sail with an energy dissipation rate ${\dot e}_{\rm diss}$. `Oumuamua  
accelerates with $a_{\rm gas } \simeq { n_{\rm ISM} m_{\rm p}  v^2 / 
\rho_{\rm O}l_3}$ until a terminal speed $v_{\rm term, gas} \sim v_{\rm O}$, 
with which ${\dot e}_{\rm gain} \simeq {\dot e}_{\rm diss}$, after it has 
traveled on a timescale $t \gtrsim \tau_{\rm{term,gas}} \sim 1$ Gyr. 
In Sect. \ref{sec:drag} we analyze the ISM gas drag effect and how it determines 
the terminal speed and causes the light sail to deviate from the intended course.

The ISM's turbulent motion is also associated with disorderly magnetic
fields, ${\vec B}$.       An electric potential drop is induced across the light sail,
and an Alfv{\'e}n wing \citep{Drell65} is launched when `Oumuamua moves with a velocity
${\vec v}_{\rm mag}$ relative to these fields. In Sect. \ref{sec:magdrag} we show that 
the dissipation along the Alfv{\'e}n wings also leads to 
a Lorentz force, which includes a drag and sideways 
acceleration. Similarly, as `Oumuamua passes through randomly 
oriented ${\vec B}$ in turbulent eddies of all sizes, $L_{\rm tur}$, the magnitude 
and direction of the magnetic drag change over time, and one can derive the 
expected value of `Oumuamua's path deviation from its initial target.

Although `Oumuamua's initial velocity does not change significantly
over shorter time intervals $\tau_{\rm{term,gas}} > t 
> \tau_{\rm rot}$ (or travel distance $L < L_{\rm{term,gas}} \sim 10$ kpc),
it endures a fluctuating sideways acceleration due to both the ISM's  ram pressure 
normal to the randomly oriented surface and magnetic drag.
Accumulation of this random-walk acceleration prompts `Oumuamua to drift
off course from its intended destination.
The expectation value of the path deviation (or impact parameter), ${\rm E}(d)
\sim ({t})^{3/2} $,
increases with $t$ and $L$ (Appendix \ref{sec:accumulative}). 

\subsection{Terminal speed and off-course deviation caused by ISM gas drag}   
\label{sec:drag}

The hypothetical tumbling light sail moving with velocity $v$ through the ISM
also experiences a drag force \citep{Laibe12}:
\begin{equation}\label{F_gas}
   F_{\rm gas} \simeq n_{\rm ISM} m_{\rm p} f_{\rm A} l_1 l_2 
   \sqrt{8\over \pi \gamma_0} c_{\rm s} v \sqrt{1+{9\pi \over 128} 
   { v^2 \over c_{\rm s}^2}}.
\end{equation}
This Epstein limit is appropriate for $l_1 \ll $ the molecular mean free path 
such that gas particles, with a mass $m_{\rm p}$, can be treated as a dilute 
medium.  {We applied the heat capacity ratio $\gamma_0 = 5/3$ for the ideal gas.} The area correction factor $f_{\rm A} 
= A/l_1 l_2 $ has a magnitude $ 1 \gtrsim f_{\rm A} \gg l_3/l_1$. 
For a light sail with $v \sim v_{\rm O} \gtrsim c_s$,   
$    F_{\rm gas} \simeq n_{\rm ISM} m_{\rm p} f_{\rm A} l_1 l_2 v^2$ with an
acceleration 
\begin{equation}   
a_{\rm gas } \simeq f_{\rm A} { n_{\rm ISM} m_{\rm p}  v^2 / \rho_{\rm O}l_3}
\label{eq:agas}
\end{equation}
and an energy dissipated rate
\begin{equation} 
{\dot e}_{\rm diss} \sim  { f_{\rm A} l_1 l_2 n_{\rm ISM} m_p v^3 } / {m_{\rm O}}.
\end{equation}
On the timescale 
\begin{equation}
\begin{aligned}
    \tau_{\rm{term,gas}} & \simeq \dfrac{\rho_{\rm O} l_3}{2f_{\rm A}n_{\rm ISM}m_{\rm p} v}  \\
    & \sim \left( 
 \dfrac{l_3/f_{\rm A} }{10^{-3} {\rm~m}} \dfrac{\rho_{\rm O}} 
{10^{3}~{\rm kg~m}^{-3}} 
\dfrac{10^{6}~{\rm m}^{-3}} {n_{\rm ISM}} 
\dfrac{10~\rm{km/s}}{ v} \right) \rm Gyr,
\end{aligned}
\end{equation}
the light sail attains an energy equilibrium with a terminal speed
\begin{equation}
\begin{aligned}
v_{\rm term,gas} &
\sim \left(\dfrac{\rho_{\rm O} l_3 {\dot e}_{\rm gain}}  
{f_{\rm A} n_{\rm ISM} m_p} \right)^{1/3} \\
& \sim 10 \left( 
\dfrac{l_3/f_{\rm A} }{10^{-3} {\rm~m}} \dfrac{\rho_{\rm O}} 
{10^{3}~{\rm kg~m}^{-3}} \dfrac{10^{6}~{\rm m}^{-3}}
{n_{\rm ISM}} \right)^{1/3} \rm{km/s}
\end{aligned}
,\end{equation}
which $\sim v_{\rm O}$ due to the drag by the volume-filling WNM on it over 
a travel distance
\begin{equation}   
L_{\rm{term,gas}} \sim \tau_{\rm{term,gas}} v_{\rm{term,gas}} \gtrsim 10 {\rm kpc}.
\end{equation}
Nevertheless, the spaceship hypothesis is based on the conjecture that the Solar 
System has always been `Oumuamua's intended destination and that it has maintained its original 
course with its initial and current $v \sim v_{\rm O}$ rather than $v_{\rm term,gas}$ despite 
their coincidentally similar magnitude.  Its corollary implies a traveled distance 
$L \leq L_{\rm{term,gas}}$ and duration $t \leq \tau_{\rm{term,gas}}$.

The ram-pressure force is 
primarily applied in the direction normal to the surface of a smooth
thin light sail rather than the direction of its motion relative to the ISM,  and
it does not sensitively depend on the ISM's turbulent speed, $v_{\rm tur}$, or eddy
size, $L_{\rm tur}$, in the limit $v \sim v_{\rm O} >v_{\rm tur}>c_s$. Moreover, due 
to the finite net torque exerted by the turbulent ISM on the light sail's asymmetric and uneven surface, which is revealed by the radiative torque measured in the case of `Oumuamua (see details in Appendix \ref{sec:ismtorque}), the spin orientation and frequency, $\omega$, of its tumbling motion  
evolve on a timescale (See Eq. \ref{eq:taurot} in Appendix \ref{sec:spinism})
\begin{equation} \label{tau_gt}
    \tau_{\rm{rot}} = {\omega \over \dot \omega} 
    \sim 1 {\left( \rho_{\rm O} l_3 \over 1 {\rm kg/m^2}  \right)} 
    {\left(n_{\rm ISM} \over 10^6 /{\rm m^3} \right)}^{-1} 
    {\left( l_1 \over 60 {\rm m}  \right)} \rm Myr < < \tau_{\rm{term,gas}}
.\end{equation}
Consequently, the ram-pressure-driven acceleration $a_{\rm drift} \sim a_{\rm gas}$ is redirected on a timescale $\tau_{\rm rot}$, which is generally smaller than the eddy turnover timescale
$\tau_{\rm tur} \sim L_{\rm tur}/v_{\rm tur}$ over a wide range of $L_{\rm tur}$ (Eq. \ref{eq:tautur}).
Over a brief travel time, ${t<\tau_{\rm rot}}$, or distance, $L=v t$, 
the light sail's spin vector does not change significantly and as such drifts with a dispersion 
velocity $\sigma_{\rm drift} \simeq a_{\rm drift} t$ in some random directions.
Over a longer travel time, $\tau_{\rm rot} \leq t < \tau_{\rm{term,gas}}$, the light sail's evolving spin 
vector and sideways acceleration, $a_{\rm drift}$, lead to a random walk in its accumulative 
$\sigma_{\rm drift} \simeq ({t / \tau_{\rm rot}})^{1/2} a_{\rm{gas}} \tau_{\rm rot} < < v$, the detail of which is shown in Appendix \ref{sec:accumulative}.   
The expected value of the deviation from the light sail's initial course for these two limiting timescales (see Eq. \ref{eq:erand} in Appendix \ref{sec:accumulative}) are
\begin{equation}
    {\rm E}(d_{\rm gas})   = \left\{
    \begin{aligned}
    & a_{\rm{gas}}t^2 / 2 ,  && {t<\tau_{\rm rot}} \\
    & ({t / \tau_{\rm rot}})^{3/2}  {a_{\rm{gas}}\tau_{\rm rot}^2 / 2} ,&& {t \geqslant \tau_{\rm rot}}.
    \end{aligned}
    \right.
    \label{eq:dgas}
\end{equation}

\subsection{{Alfv{\'e}n-wing drag by turbulent WNM}}
\label{sec:magdrag}
The ISM also contains magnetic fields with $B \approx 10 \,\rm \mu G = 1 \rm \,nT $
in the neutral medium and a few times larger in the molecular clouds \citep{Draine11}. 
As a light sail traverses the magnetic field with a relative velocity ${\vec v}_{\rm mag}$, an electric field 
${\vec v}_{\rm mag} {\bf \times} \vec B$ is induced with a potential drop $U  \simeq \langle l \rangle 
v_{\rm mag} B$ across the time-averaged length, $\langle l \rangle$, of the tumbling light sail. 
An Alfv{\'e}n wave is emitted along a wing \citep{Drell65}, leading to a drag against ${\vec v}_{\rm mag}$ with an acceleration 
\begin{equation}  
a_{\rm{mag}} \simeq 
{f_{\rm D} B^2 v_{\rm mag} / 2\rho_{\rm O} l_3 \mu_0 \bar v_A}
\label{eq:amag}
\end{equation}
in the direction $( {\vec B} \cdot {\vec v}_{\rm mag}
{\vec B} - \vert {\vec B} \vert^2 {\vec v}_{\rm mag})/\vert {\vec B} 
\vert ^2 \vert {\vec v}_{\rm mag} \vert$,  where $f_{\rm D} \simeq 
\langle l \rangle / l_1$, with $1 \gtrsim f_{\rm D} > > l_3/l_1$, $\mu_0 (= 4 \pi 
\times 10^{-7} \rm{N \, A^{-2}})$, $v_A = B/\sqrt{\mu_0 \rho_g}$ is the Alfv{\`e}n speed, 
$\bar v_A = \mu_0 v_A \sqrt{1+M_A^2 + 2M_A  \cos \gamma } \sim 10-26 \,\rm{km/s}$ 
in the diffuse ISM, $M_A = v_{\rm mag}/v_A$ is the magnetic Mach number, 
and the angle $\cos \gamma = { {\vec v}_{\rm mag} \cdot \vec B / (|\vec v_{\rm mag}| \cdot |\vec B|)}$.  
When the rate of energy dissipation per unit mass in the Alfv{\' e}n wing, 
\begin{equation}   
    {\dot e}_{\rm mag} \simeq {f_{\rm D}l_1^2 v_{\rm mag}^2 B^2}/{\mu_0 {\bar v}_{\rm A} m_{\rm O}} 
,
\end{equation}
is balanced by ${\dot e}_{\rm gain} \simeq v_d^2/2 \tau_\ast$, a terminal speed is established with
\begin{equation}
    \begin{aligned}
    v_{\rm term,mag} & \simeq 
\left( 
\dfrac{ \rho_{\rm O} l_3 \mu_0 \bar v_{\rm A} }  {2f_D B^2 \tau_\ast}
\right)^{1/2} v_{d} \\
& \sim 7.4 \left( 
 \frac{l_3/f_{\rm D} }{10^{-3} {\rm~m}} \dfrac{\rho_{\rm O}} 
{10^{3}~{\rm kg~m}^{-3}} \dfrac{\bar v_{\rm A}}{10~\rm{km/s}} \right)^{1/2} 
\left ( \dfrac{1{\rm~nT}}{B} \right) \rm{km/s}        
    \end{aligned}
,\end{equation}
which $\sim v_{\rm O}$ with $f_{\rm D} \sim {\mathcal O} (1)$.
The timescale to reach $v_{\rm term,mag}$ is
\begin{equation}
\begin{aligned}
     \tau_{\rm{term,mag}} &\simeq \dfrac{\rho_{\rm O} l_3 \mu_0 \bar{v_{\rm A}}}{2f_{\rm D} B^2}  \\
     & \sim 0.55 \left( 
 \dfrac{l_3/f_{\rm D} }{10^{-3} {\rm~m}} \dfrac{\rho_{\rm O}} 
{10^{3}~{\rm kg~m}^{-3}} \dfrac{\bar v_{\rm A}}{10~\rm{km/s}} \right) 
\left ( \dfrac{1{\rm~nT}}{B} \right)^2 \rm{~Gyr}    
\end{aligned}
\end{equation}
over a travel distance 
\begin{equation}
    L_{\rm{term,mag}} \sim \tau_{\rm{term,mag}} v_{\rm{term,mag}} \gtrsim 4 {\rm kpc}.
\end{equation}
Based on the same assumption that the light sail has preserved its original course,
we only considered the limit $L \leq L_{\rm{term,mag}}$ and $v \simeq v_{\rm O}$. 

The magnetic fields in the diffuse ISM are characterized by chaotic structure, although they are more orderly 
in star-forming GMCs \citep{Crutcher2012}. We assume they are coupled to the large eddies in the WNM, which 
has a turbulent speed $v_{\rm tur} \sim (L_{\rm tur}/ {\rm pc})^{0.38}$ km s$^{-1}$ 
that persists on the timescale \citep{Choudhuri19}
\begin{equation} 
\tau_{\rm tur}= L_{\rm tur}/v_{\rm tur} \sim (L_{\rm tur}/ 
{\rm pc})^{0.62} {\rm  Myr} < < \tau_{\rm{term, mag}}
\label{eq:tautur}
\end{equation} 
on size scale $10^{-2} {\rm pc} \lesssim L_{\rm tur} 
\lesssim 10^2 {\rm pc}$. 
The orientation between `Oumuamua $\vec v_{\rm mag}$ and the background
$\vec B$ (or equivalently $\gamma$) evolves on its travel timescale
across eddies on scales with measured turbulent fields ($L_{\rm tur} \lesssim
1$ pc), $\tau_{\rm tur} \sim L_{\rm tur}/v_{\rm O} \sim 0.1$ Myr. 
This revision leads to greater variations, over a shorter timescale but on the 
same turbulent-eddy length scale, in the direction of the drag 
force on `Oumuamua than those due to the limited modification in the 
amplitude $v_{\rm mag}$ (Eq. \ref{eq:amag} and Fig. \ref{fig:solarsail_rotation}).

For the spacecraft hypothesis, the initial value of $v$ is preserved ($\sim v_{\rm O}$) 
over a travel distance $L \leq L_{\rm{term,mag}}$. 
Over a brief travel time interval, $t<\tau_{\rm tur}$, the direction and magnitude of the drag
by the turbulent fields on the light sail is sustained such that  $a_{\rm drift} 
\simeq a_{\rm{mag}}$ and dispersion velocity $\sigma_{\rm drift} \simeq a_{\rm mag} t$.
Over a longer duration with $\tau_{\rm tur} \leq t \leq \tau_{\rm{term,mag}}$, the 
light sail encounters a new turbulent eddy with different $v_{\rm tur}$, $\gamma$, and  
$L_{\rm tur}$.  While the magnitude of its $a_{\rm drift} \sim a_{\rm{mag}}$, random
walk in the direction of relative motion leads to accumulative $\sigma_{\rm drift} \simeq 
({t / \tau_{\rm tur}})^{1/2} a_{\rm{mag}} \tau_{\rm tur}$.
The expected values of the deviation from the light sail's initial course for these two limiting timescales are\begin{equation}
    {\rm E}(d_{\rm mag})   = \left\{
    \begin{aligned}
    & a_{\rm{mag}}t^2 / 2 ,  && {t<\tau_{\rm tur}} \\
    & ({t / \tau_{\rm tur}})^{3/2}  {a_{\rm{mag}}\tau_{\rm tur}^2 / 2} ,&& {t \geqslant \tau_{\rm tur}}
    \end{aligned}
    \right.
    \label{eq:dmag}
.\end{equation} 
Since the magnitude of $v_{\rm tur}$ and $\tau_{\rm tur}$ are increasing functions of $L_{\rm tur}$, that of 
$a_{\rm mag}$ and ${\rm E}(d_{\rm mag}) $ also increases with $L_{\rm tur}$ (Eqs. \ref{eq:amag} and \ref{eq:dmag}).
In the present context, we estimated a lower limit for ${\rm E}(d_{\rm mag})$  by assuming the small $L_{\rm tur}
\lesssim 1$ pc, where turbulent fields can be resolved.  We also neglected the $L_{\rm tur}$ dependence in $B$.

The total trajectory deflection has an expected value ${\rm E}(d_{\rm{ISM}}) 
= \sqrt{{\rm E}(d_{\rm gas}^2) + {\rm E}(d_{\rm mag}^2) }$ with a Rayleigh distribution (See Eq. \ref{eq:rayleigh} in Appendix \ref{sec:accumulative}).  
In the diffuse WNM, ${\rm E}( d_{\rm gas} ) < {\rm E}( d_{\rm mag} ) $, $\tau_{\rm tur} \sim 0.1$ Myr such that 
${\rm E}(d_{\rm{ISM}}) \sim {\rm E}(d_{\rm mag})$ and the light sail's sideways motion undergo Levy walk over 
a travel length scale $L \gtrsim 1$ pc. In more dense CNM, ${\rm E}(d_{\rm{ISM}}) \sim
{\rm E}( d_{\rm gas} ) \sim {\rm E}( d_{\rm mag} )$, $\tau_{\rm rot} \sim 0.1$ Myr and the sideways drift 
also follows a Levy walk path over travel length scale $L \gtrsim 1$ pc.  Although ${\rm E}(d_{\rm{ISM}}) \sim
{\rm E}( d_{\rm gas} ) > {\rm E}( d_{\rm mag} )$ in relatively dense ($n_{\rm ISM} \gtrsim 10^{10}$ cm$^{-3}$) GMCs, 
their hydrodynamic drag may lead to limited drift, despite their short $\tau_{\rm rot}$ and $\tau_{\rm term, gas}$,
due to their sparse volume filling factor.  

In the evaluation of $a_{\rm mag}$, we neglected the light sail's spin because $v_{\rm O} > > 2 \pi l_1/P$.
Nevertheless, this motion also leads to the excitation of Alfv{\'e}n waves at both ends and dissipation
of its spin energy $E_{\rm rot} = {I \omega^2/ 2 }$ \citep{Zhang20} with a power
\begin{equation}
    P_{\rm rot} = {B^2 \omega^2 l_1^4 \sin^2 \gamma \over 4 \mu_0 v_A}.
\end{equation}
The damping timescale of rotation is
\begin{equation}
    \tau_{\rm{rot,mag}} = {E_{\rm rot} \over P_{\rm rot}} =  { \rho_{\rm O} l_3 \mu_0 v_A  \over B^2  \sin^2 \epsilon } \sim 20 {\left( \rho_{\rm O} l_3 \over \rm 1 kg/m^2 \right)} \rm Myr
\end{equation}
if the Alfv{\'e}n speed is assumed to be $1 \, \rm km/s$ in interstellar space. 
We see that this damping timescale is much longer than that due to the ISM's ram
pressure, $\tau_{\rm{rot}}$ (Eq. \ref{eq:taurot}), and eddy turnover timescale, 
$\tau_{\rm tur}$. In the determination of relative 
velocity between it and the magnetic field, the light sail's rotational 
damping due to magnetic torque can be neglected. Overall, we show that the 
rotational evolution timescale is inversely proportional to the gas 
number density and $\tau_{\rm{tur}} \approx 0.1\tau_{\rm{rot}} \approx  0.1 \, \rm Myr$ for $n_{\rm ISM} \sim 10^6 \, \rm m^{-3}$. The deflections induced by the magnetic field and the gas pressure are comparable (${\rm E}(d_{\rm mag}) \approx {\rm E}(d_{\rm gas})$).

Under the assumption that a light sail's original $v$ is preserved before it enters the Solar System, 
there is no a priori reason for its $v$ to be correlated with its $l_1$ or $l_2$.  In contrast, the distance traveled, $L$, by freely floating interstellar asteroids and comets (with $l_3$ not much smaller than $l_1$ or $l_2$) is likely to be greater than min($L_{\rm{term,gas}}, L_{\rm{term,mag}}$), and as such their motion will be gravitationally accelerated, dragged by 
the interstellar gas, and retarded by the interstellar magnetic field through Alfv{\'e}n-wing drag \citep{Zhang20}.
When a dynamical equilibrium is established by the balance of these effects, the ISOs attain a 
terminal speed relative to the LSR, $v_{\rm d} (l_1) \simeq$ min($v_{\rm term, gas}, 
v_{\rm term, mag}$), which depends on their sizes.

\section{Dynamics of a tumbling light sail in the Solar System}
\label{sec:3}

\subsection{Sideways radiation force}

During its sojourn through the Solar System, `Oumuamua endures intense heating near its perihelion.
The surface temperature of the light sail increases rapidly, resulting in an increased average atomic 
kinetic energy of material. The atoms oscillate with anharmonic potentials in greater amplitudes, and 
therefore the material tends to expand with a larger separation between atoms. The fractional change 
in the light sail's surface area can be estimated as
$    {\Delta S / S } = \alpha_A \Delta T \times 100 \% $,
where $\alpha_A$ is the thermal expansion coefficient. Molaro et al (2019) estimate $\alpha_A = 8 \times 10^{-6}$ 
for boulders and $\alpha_A = 2.4\times 10^{-4}$ for regolith. The thermal expansion coefficient, $\alpha$, changes 
over the temperature as
\begin{equation}
    \Delta \alpha_A \approx 4\times 10^{-9} \Delta T.
\end{equation}
The thermal equilibrium temperature at $p_{\rm O}$ is $\sim 800$ K, so the thermal expansion coefficient, $\alpha$, 
is nearly constant. The expansion ratio of the area is $\Delta S / S \approx 16\%$ for regolith material and 
$\Delta S / S \approx 0.6\%$ for boulder material.  Such thermal stress would not significantly alter the 
shape or surface roughness of the light sail.

However, the solar radiation pressure exerts a sideways force on `Oumuamua analogous to 
that due to ram-pressure drag (Fig. \ref{fig:solarsail_rotation}). When photons hit the surface, 
some are reflected immediately while some get absorbed and reradiated with a time lag. The latter effect 
is negligible since the thickness we consider here is so thin that it could be smaller than the penetration depth 
of the thermal wave, which leads to a nearly equal temperature all over the surface.
The immediately reflected photons cause a recoil force $F \simeq \Phi A/c$ in
the direction perpendicular to the surface.  Reflection on its rough surface also leads to a 
torque $T_{\rm rad}$ (Eq. \ref{eq:radtorque}), which cause `Oumuamua's spin period, $P_{\rm O}$, 
and orientation to evolve on a timescale $\tau_{\rm rad} > > P_{\rm O}$ (Eq. \ref{eq:taurotrad}).

The direction of this mean radiative force (integrated over a precession period) is 
about the wobble angle, $\theta $, of `Oumuamua and the angle between the incident light and `Oumuamua’s angular 
momentum, $\theta^\prime$. The heliocentric radial and sideways forces (in the antisolar and transverse directions) are
\begin{align}
    & \vec F_{\rm mean,r} = (\vec F_{\rm mean} \cdot \vec e_r) \vec e_r \\
    & \vec F_{\rm mean,s} = \vec F_{\rm mean} - \vec F_{\rm mean,r} .
\end{align}

The deflection angle of the force can be defined as
\begin{equation}
    \alpha = \arccos\left( {F_{\rm mean,r} \over F_{\rm mean}} \right).
\end{equation}

The values of the angle, $\alpha$, for different $\theta$ and $\theta^\prime$ are shown 
in Fig. \ref{fig: force_ratio}. 
In the trajectory of `Oumuamua, the value of $\theta^\prime$ varies. Since information on `Oumuamua's spin (e.g., 
the directions of its spin angular momentum vector, principle axis, and precession angle) is not available, we used the 
mean value of $\alpha$ as an approximation. The mean value is $\bar \alpha \approx 26^\circ$, which gives the ratio 
of the sideways force to the radial force
\begin{equation}
    {F_{\rm mean,s} \over F_{\rm mean,r}} = \tan \bar \alpha \approx {1\over 2}.
\label{eq:fradsidrad}
\end{equation}
We can simply use this value to estimate the sideways deflection of `Oumuamua’s orbital paths through the Solar System.

Since the position of `Oumuamua is poorly constrained, we applied a Monte Carlo simulation to `Oumuamua's orbital evolution over 80 days by testing $10^5$ cases with random initial orientations of angular momentum. Since the observation timescale of de-spin is $\sim 24$ days, we reset the angular momentum to a random orientation every 24 days. For each case, we calculated the mean ratio of the sideways force to the radial force along the way. The sideways force will lead to a displacement that is normal to the Keplerian orbital plane. The comparison of the result of the light sail and `Oumuamua is shown in Sect. \ref{sec:displacement}.

In addition, the nongravitational model that fits the observation data best indicates that the radial force is 
$A_1 r^{-2}$, with $A_1$ being constant. However, the  tumbling motion of the light sail also leads to a variation 
in the magnitude of the force due to the time-varying orientation of the thin sheet. The timescale of the de-spin 
is $\sim 24 \, \rm day$, which results in considerable changes in the angular momentum vector. We have already shown 
that the radiation force relies on the angular momentum (Eq. \ref{eq:fmeanbeta}), so we would expect 
a large variation range of the radiation force during 80 days of observation instead of a constant coefficient 
$A_1$. At the very least, even if the angular momentum does not change a lot during the orbit, the light angle 
($\theta^\prime$) changes due to the orbital motion, which also leads to a time-varying radiative radial force. 
During the observational period (80 days), the mean-ratio changes of the radial force, $\Delta A_1/A_1$, 
estimated from $10^4$ randomly generated test particles, show 
a normal distribution. The standard deviation is $\sim 0.3,$ which indicates that the coefficient of the nongravitational force is expected to fluctuate by $\sim 30\%,$ while its reported value of 
$A_1 = 4.92 \pm 0.16  \times 10^{-6} \, \rm m/s^2$ is inferred \citep{Micheli18} 
under the assumption of no change (i.e., $\Delta A_1=0$).

Since the light curve and the radiation force components highly depend on the orientation of the light sail, they build a correlation with each other that can be examined for ISOs in the future. The correlation can be described quantitatively by the circular cross-correlation function:
\begin{equation}
    R_{H,a}(\tau) = {\sum_{i = 0}^{N-1}  H_i' a_{I+\tau}^\prime\over (\sum_{i = 0}^{N-1}  H_i'^2)^{1/2}  (\sum_{i = 0}^{N-1}  a_i'^2)^{1/2}},
\end{equation}
where $H_i^\prime$ and $a_{I+\tau}'$ are the i-th and $(i+\tau)$-th elements in the time-sequence absolute magnitude signal, $H'$, and the acceleration, $a'$, during a rotation cycle, which is centralized by
\begin{align}
    H^\prime = H - \bar H, \\
    a' = a - \bar a.
\end{align}
Here $\bar H$ and $\bar a$ are the mean values of $H$ and $a$. Considering the periodicity of the light curve and the radiation force, we made a periodic extension of $H'$ and $a'$, which leads to a modified correlation function:
\begin{equation}
    R_{H,a}'(\tau) = {\sum_{i = 0}^{N-1}  H_i' a_{\rm mod (i+\tau, N)}'\over (\sum_{i = 0}^{N-1}  H_i'^2)^{1/2}  (\sum_{i = 0}^{N-1}  a_i'^2)^{1/2}}
.\end{equation}
Therefore, the correlation function is a periodic function with the same period as $H’$ and $a’$ and we only need to pay attention to its behavior in one period. The correlation function reveals the similarity of the two signals at a time delay $\tau$. The value range of $R_{H,a}'(\tau)$ is [-1,1], and a larger value of $\vert R_{H,a}'(\tau) \vert$ implies a higher correlation between $H$ and $a$.

\subsection{{Light curve and visibility}}
\label{sec:light_curve}

\begin{figure*}
    \centering
    \includegraphics[width=\linewidth]{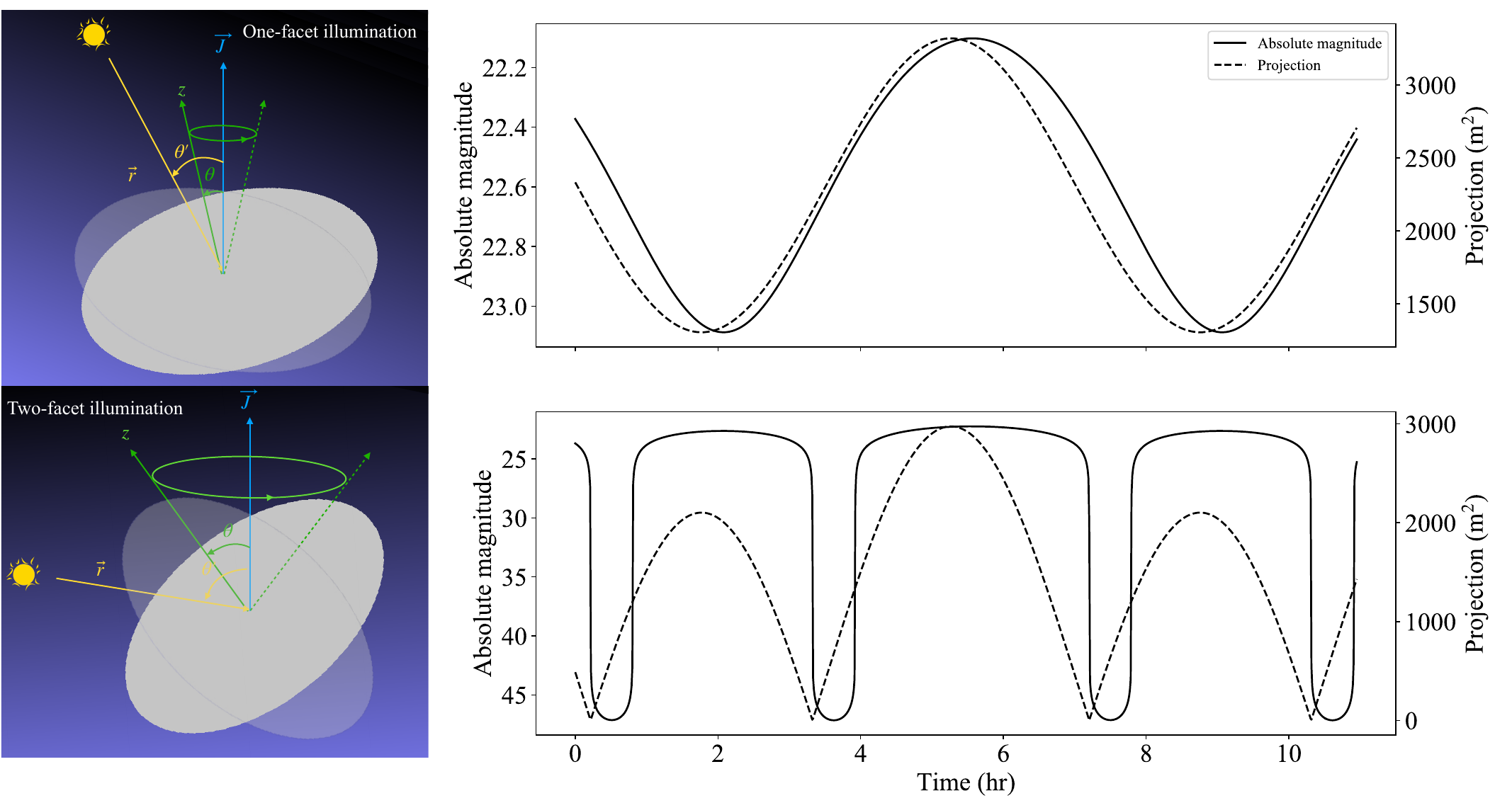}
    \caption{ {Illustration of the light sails in the one-facet illumination state and two-facet illumination (left panels)} and their light curves (right panels).  {The rotation angular momentum is $\vec J,$ and the spin vector is along the major principal axis, $z$.} The position vector $\vec r$ is pointing `Oumuamua away from the Sun. The solid lines denote the absolute magnitude, while the dashed line denotes the areas of projection along the sunlight direction. The imperfect consistency results from the nonzero phase angle  {$\sim 20^\circ$ \citep{Fraser18}}. }
    \label{fig:lightcurve of light sail}
\end{figure*}

The light curve of a tumbling light sail can exhibit a very large amplitude due to the large dimensional ratio. For a light sail with three axes of $60  \rm m \times 60 \,\rm m \times 0.3 \, \rm mm $, the amplitude could reach as large as $2.5 \log (60 \,\rm m/0.3 \,\rm mm) \sim 13.25$  {in the case that} the phase angle of the light sail is equal to $0^\circ$, which means the light sail is on the opposite side of the Sun relative to Earth. With a nonzero phase angle, the amplitude is even larger \citep{Fraser18}. We used a symmetric ellipsoid ($l_1 = l_2$) model with an an isotropic-scattering surface for the calculation of the absolute magnitude of the light sail. The absolute magnitude is expressed as \citep{Mashchenko19,Muinonen15}
\begin{equation}
\begin{aligned}
    H &= \Delta V - 2.5 \log
    l_3' {T_{\oplus} T_\odot \over T}  - 2.5\log [\cos(\lambda' - \alpha') \\
    & + \cos \lambda' + \sin \lambda'  \sin (\lambda' - \alpha') \ln \left( \cot { \lambda'\over 2} \cot {1\over 2} (\alpha' - \lambda')\right) 
     ] ,  
\end{aligned}
\end{equation}
where
\begin{align}
    &\vec T_{\odot} = (S_1, S_2,  S_3 /l_3', ), \\
    &\vec T_{\oplus} = (E_1,E_2, E_3 /l_3' ), \\
    &\vec T = \vec T_{\odot} +\vec T_{\oplus} .  
\end{align}
Here $\vec S = (S_1, S_2,S_3)$ and $\vec E = (E_1, E_2,E_3)$ are the position vectors of the Sun and Earth in the corotating coordinate system. 
The parameter $l_3'$ is defined as $l_3' = l_3 /l_1 $
The angle $\alpha'$ is the angle between $\vec T_{\odot} $ and $\vec T_{\oplus}$, 
and $\lambda'$ is the angle between $\vec T_{\odot} $ and $\vec T$. The parameter $\Delta V$ is the offset between the model and the observed light curve.

\begin{figure*}
    \centering
    \includegraphics[width=\linewidth]{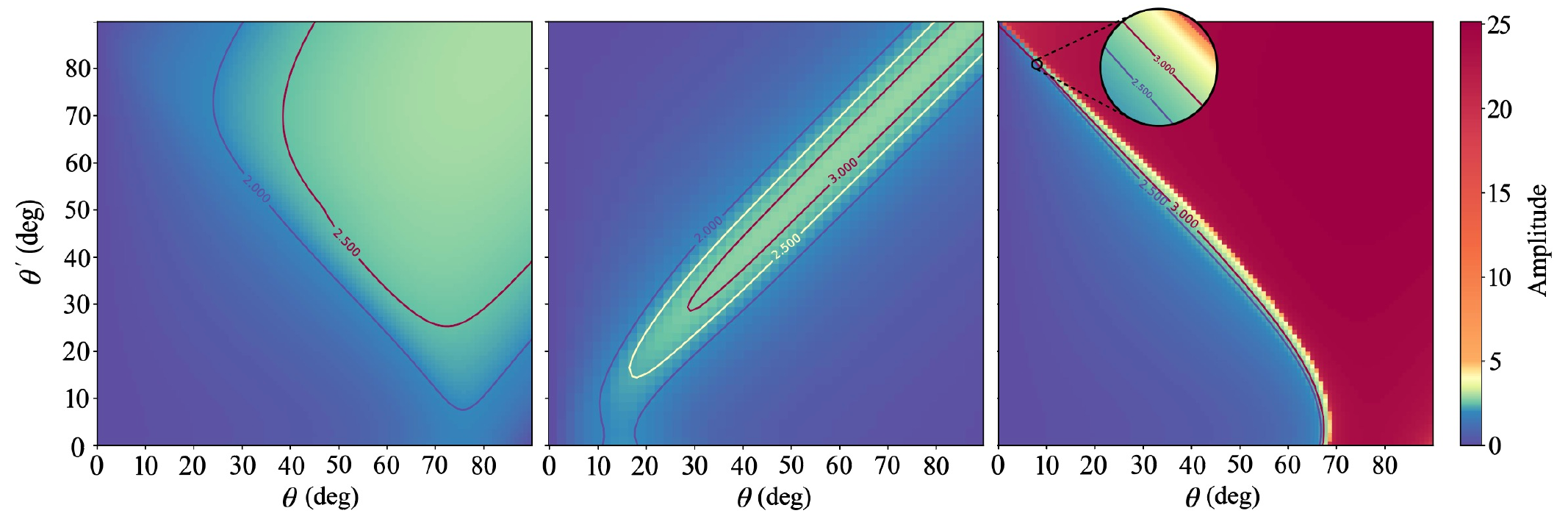}
    \caption{Amplitude map for the light curves of an oblate shape (left),  prolate shape (middle), and light sail (right). The fractions of area where the amplitude $\Delta H$ is between 2.5 and 3 are 0.34, 0.1, and 0.015 for the oblate shape, prolate shape, and light sail, respectively.}
    \label{fig:distribution of amplitude}
\end{figure*}

As already mentioned, the light curve can show an extremely large amplitude, which is shown in Fig. \ref{fig:lightcurve of light sail}. Whether the amplitude is exotically large depends on the rotation state of the light sail and its orientation to the Sun, which can be categorized into two types.
The first is the one-facet illumination state (OIS), where the light sail is in such a rotation that only one facet is exposed to sunlight. In this case, the projected area of the sunlight direction cannot reach the minimum value ($l_1 \times l_3$) except when the sunlight comes from the edge of the light sail. Therefore, in the OIS the amplitude is usually not exotically large.

The second is the two-facet illumination state (TIS), where two facets of the light sail are successively exposed
to sunlight due to rotation. In this case, the projected area of the sunlight direction reaches the lowest point the moment 
when the illuminated facet transfers from one to the other. In the TIS, the light curve of the light sail exhibits an extremely large
amplitude, and the light sail becomes invisible for some time in its rotational cycle.

Figure \ref{fig:lightcurve of light sail} shows two examples of the light curve of the light sail, one for the OIS and 
the other for the TIS. Due to the symmetry of the square light sail, the periodical variation in
the orientation of the light sail can be described by one parameter, the wobble angle ($\theta),$ 
as Fig. \ref{fig:lightcurve of light sail} shows. Another variable that affects the light curve is 
the direction of the position vector of the light sail relative to the Sun, which is equally 
translated to the illumination angle, $\theta^\prime$. The light sail is in the OIS when 
$\theta^\prime+ \theta \leq 90^\circ$ and is otherwise in the TIS.

Figure \ref{fig:distribution of amplitude} shows the amplitude $\Delta H$ maps over a ($\theta$, $\theta^\prime$) parameter space for the oblate shape, the prolate shape, and the light sail. We can see a clear boundary between the OIS and the TIS of the light sail in the amplitude map, although it does not strictly follows $\theta^\prime+ \theta \leq 90^\circ$ due to the nonzero phase angle.

\section{Comparison between a free-floating light sail and `Oumuamua}
\label{sec:4}

\subsection{{Probability of close perihelion passage and occurrence timescale}}
\label{sec:closeencounter}

\begin{figure*}
    \centering
    \includegraphics[width=\linewidth]{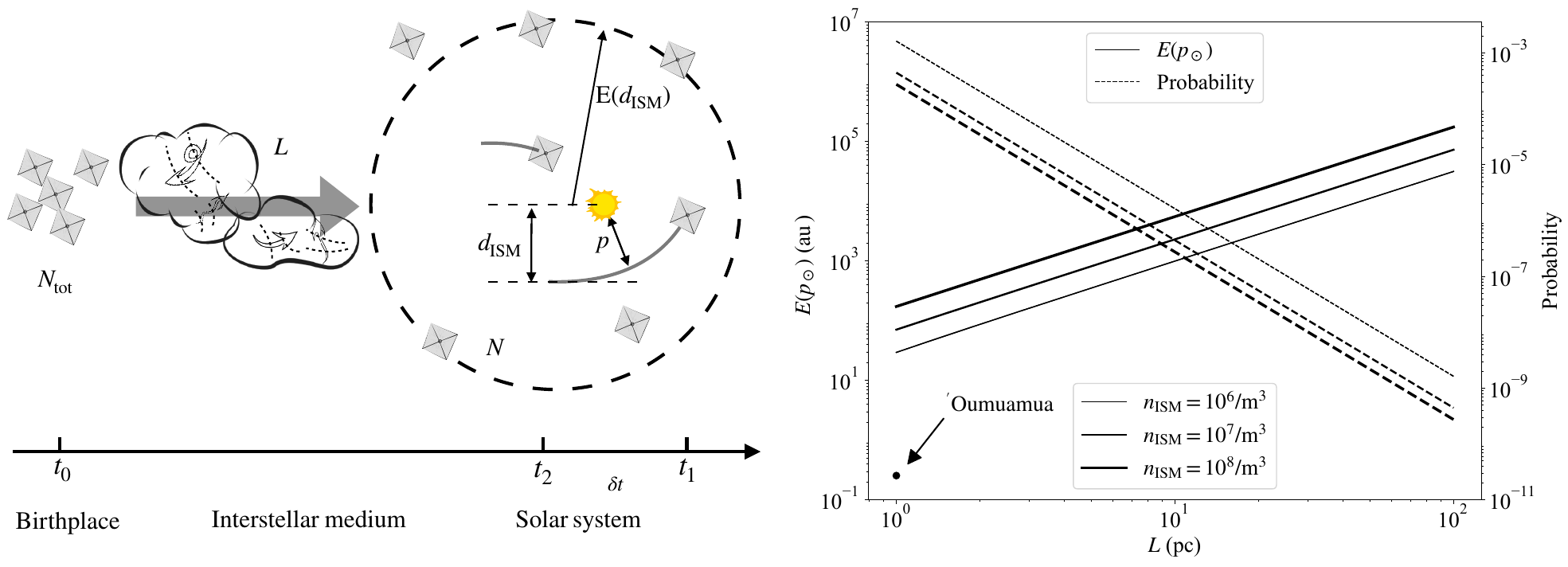}
    \caption{Schematic illustration of the diverse pathways of a constellation of $N_{\rm tot}$ 
    tumbling solar sails traveling over a distance $L$ through the turbulent magnetized ISM (left panel) and the expected value of the perihelion distance as a function of travel distance (right panel). As shown in the left panel, dispersive 
    hydrodynamic and magnetic drag lead to a sideways drift, $d_{\rm ISM}$, and a dispersion in time of
    arrival, $\delta t$, at  the Sun. The expected value ${\rm E}(d_{\rm ISM})$ (dashed gray circle) 
    is generally much  larger than the targeted `Oumuamua's perihelion distance, $p_{\rm O}$. In the right panel, the expected value of the perihelion distance, ${\rm E}(p_\odot)$ (solid lines), is shown as 
    a consequence of deflection after traveling over a distance $L$ through the ISM with a number 
    density $n_{\rm ISM}$.  {The probability of a single light sail striking the Solar System at a distance $p_\odot \leq p_{\rm O}$ (represented by a solid dot) from the Sun is shown with dashed lines.}}
    \label{fig:travel}
\end{figure*}

In Sects. \ref{sec:drag} and \ref{sec:magdrag} we model the dynamics of `Oumuamua 
as a free-floating light sail traveling through a turbulent magnetized ISM under the influences 
of gas and magnetic drags on an intended journey to the Solar System. 
Together, these drag forces lead to a deviation with a Rayleigh distribution
around an expected value ${\rm E}(d_{\rm{ISM}}) \simeq \sqrt{{\rm E}(d_{\rm gas}^2)  
+ {\rm E}(d_{\rm mag}^2) }$,  which is equivalent to an impact parameter in 
the gravitational domain of the Sun. The corresponding expected value 
of the perihelion distance $p_\odot = d_{\rm ISM} \left( \sqrt{ \xi^2 + 1} 
- \xi \right)  \simeq d_{\rm ISM} $,  where $\xi = G M/v^2 d_{\rm ISM} < < 1$ 
and $p_\odot$ is generally much larger than `Oumuamua's observed $p_{\rm O}$ 
(Fig. \ref{fig:travel}).

The probability of `Oumuamua attaining $p_\odot$ smaller than  {`Oumuamua's} observed perihelion $p_{\rm O} (0.26$ au) is
\begin{equation}
    P(p_\odot < p_{\rm O}) = 1-\exp \left(-{p_{\rm O}^2  \over (t/\tau_{\rm tur})^{3} \Delta d^2 } \right)
    \label{eq:periprobability}
,\end{equation}
where $\Delta d = a_{\rm ISM} t_{\rm tur}^2/2$.  For various values of $n_{\rm ISM}$, $P(p_\odot < p_{\rm O})$
is smaller than $10^{-3}$ over a travel distance $L \gtrsim 1$ pc, and it scales as $P \propto L^{-3}$ for larger
$L,$ which is shown in Fig. \ref{fig:travel}.

\begin{figure}
    \centering
    \includegraphics[width=\linewidth]{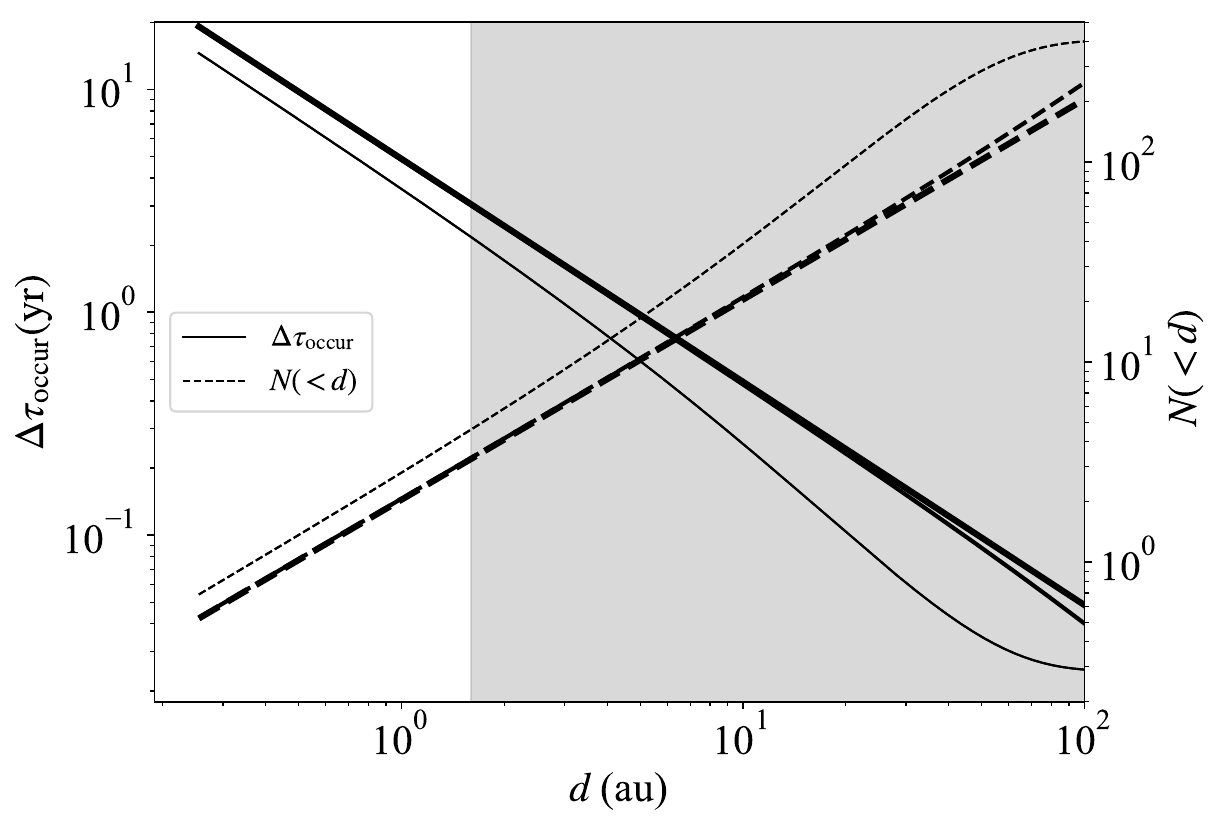}
    \caption{Occurrence timescale (solid lines) and cumulative number (dashed lines) of `Oumuamua's cohorts within $d$ of the Sun for travel distances in the ISM $L = 1\, \rm pc,\, 10\,pc,\, 100\, pc$ (increasing line widths).  {The timescale of occurrence of a light sail grows with increasing  traveling distance, $L$, as a result of diffusion in the interstellar space. The timescale of occurrence within 100 au is insensitive to the travel distance, $L$, when $L >10~\rm pc$.} The unshaded region is the observational domain of the first $\sim 10$ years of the Vera Rubin telescope.}
    \label{fig:number_occurrence}
\end{figure}

One possible solution for reconciling the low close-encounter probability with the expeditious discovery of `Oumuamua
is the assumption that it was dispatched at one time with a population of $N_{\rm tot}$ cohorts, among which it had 
the smallest $p_\odot$.  Based on the $N_{\rm obs} =1$ observation of `Oumuamua with $p_{\rm O} = 0.255$ au, we infer
$N_{\rm tot} = {N_{\rm obs} / P(p_\odot<p_{\rm O})}$ as a function of $d$.
If they were launched concurrently at the same location, they would diffuse with a nearly uniform density 
$n = {N_{\rm tot} / d_{\rm ISM}^3}$  within a patch with a size $\sim d_{\rm ISM}$ (Fig. \ref{fig:travel}).  This dispersion 
would lead to a spread in the arrival time near the Sun $\tau_{\rm occurrence} = 2d_{\rm ISM}/v_{\rm O} (n -1)$, where $n$ is the number per unit distance to the Sun (\#/au) given by $N_{\rm tot} p(d)$. Figure \ref{fig:number_occurrence} shows that 
`Oumuamua-like objects would occur more frequently in the outer Solar System. A shorter distance for `Oumuamua's birthplace leads to a shorter occurrence timescale. However, none of `Oumuamua's hypothetical cohorts has been discovered yet, implying that `Oumuamua comes from a distant place where it has a much lower probability of reaching the Solar System (Fig. \ref{fig:travel}). 

With its limiting sensitivity ({24} magnitude),  {the 
Vera Rubin telescope can detect an `Oumuamua-like object out to $\sim 1.6$ au,
and the number of cohorts passing through this distance is 
extrapolated to occur over time  {at} intervals $\Delta \tau_{\rm occur} \sim 2-3$ yr, which are
insensitive to the distance they have traveled through the ISM (Fig. \ref{fig:number_occurrence}).  
The accumulated discoveries over the telescope's first 10 years of 
operation \citep{Abell2009} {correspond} to a significant fraction of $N (d \leq  1.6 \,\rm au) \sim 3-5 $
independent cohorts, {from the same direction in interstellar space},
which is smaller than that estimated for freely floating ISOs ($\sim 10$)\footnote{https://lsst-sssc.github.io/sciCases.html}.} Although this 
random drift may be rectified with a sophisticated guiding system, such a remote-control
prowess during its arduous journey would be difficult to reconcile with `Oumuamua's apparently 
freely tumbling motion, inferred from its evolving light curve (with ${\dot P}_{\rm O}$) and small
sideways displacement, relative to that in the antisolar direction (Sect. \ref{sec:displacement}) without introducing an additional assumption regarding an accidental guiding-system failure 
during its most recent Solar System passage.

\subsection{Variability and visibility of `Oumuamua's light curve}

After an initial flurry of high-cadence observations (mostly provided by discretionary-time allocation), regular
follow-up monitoring of `Oumuamua continued for another two months.  Based on this limited data set, the following 
useful information can be extracted. 

First, the amplitude of `Oumuamua's light curve on  {October 25, 2017,} from {the Very Large Telescope (VLT)} data is $\Delta H \sim$ 2.5 magnitudes, and the 
that on  {October 27, 2017,} from {the Canada–France–Hawaii Telescope (CFHT)} and {the United Kingdom Infra-Red Telescope (UKIRT)} data is $\Delta H \sim$ 3 magnitudes \citep{Meech17}. The true amplitude could 
be larger due to the large error on the faintest point in the light curve, and thus it is safe to conclude that the 
amplitude was in range $\Delta H \simeq$ (2.5, 3) from  {October 25, 2017,} to October 27, {2017}.

Second, `Oumuamua is visible in later observations, from  {November 15, 2017, to January 2, 2018}, which means it is at least brighter than 27 magnitudes \citep{Micheli18}.

Based on the above data, we investigated the probability of `Oumuamua being a light sail by considering the probabilities of the following
two independent characteristics.
(1) Regarding the {variability,} the light curve amplitude of `Oumuamua as a light sail is in the range $\Delta H \simeq$ (2.5, 3), based on the first fact. (2) Regarding the {visibility}, `Oumuamua as a light sail is brighter than 27 magnitudes and so is visible on each observational attempt, 
based on the second fact.

\begin{figure}[htbp]
\centering
\includegraphics[width=\linewidth]{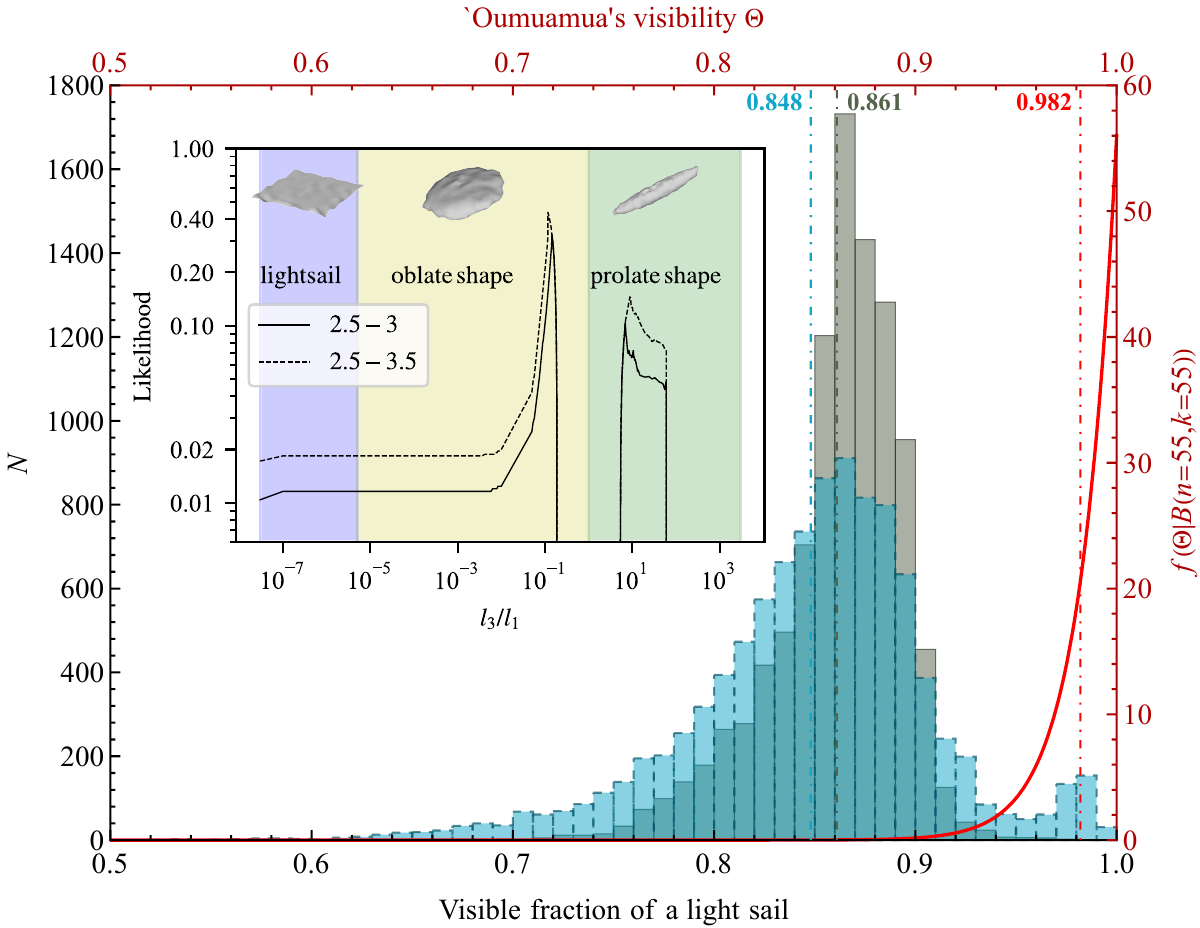}
\caption{Histogram of $10^4$ simulations of visible probability of a light sail versus posterior visible probability density distribution of `Oumuamua over 49 days. The gray histogram represents the statistical result from a uniform random sampling, whereas the blue histogram is the statistical outcome from a nonuniform sampling that mimics ground-based and HST astrometry of `Oumuamua. The dot-dashed gray and blue lines 
denote the mean values of these two statistical visibility simulations of a light sail, respectively.  {The dot-dashed and solid red lines represent the expected a posteriori estimation and the posterior probability distribution of `Oumuamua's 
visible probability (see Eq. \ref{eq:posterior}), respectively, based on the fact that `Oumuamua was visible in all 55 observations.}   The embedded panel shows the likelihood that the amplitude of the light curve 
varies within the ranges $\Delta H \simeq$ (2.5, 3) \citep{Meech17} and (2.5, 3.5) during the period 
  {October 25-27, 2017,} for the light-sail, oblate shape, and prolate shape models.}
\label{fig:visibility}
\end{figure}

To calculate the probability of event A, we tested $10^4$ models, with the variable parameters wobble angle, $\theta$, and the illumination angle, $\theta^\prime$, each sampled with $10^2$ values uniformly distributed over ($0^\circ, 90^\circ$). Therefore, each of the $10^4$ models has an equal probability. Counting the number of models where the amplitude is in range $\Delta H \simeq$ (2.5, 3) and dividing this number by $10^4$, we obtain the probability of event A. The likelihood function of the dimension ratio $f_l$ given event A is expressed as 
\begin{equation}
    L (f_l | {\rm A}) = P_{f_l}(\rm{A}).
\end{equation}
Therefore, applying the maximum likelihood estimation (MLE), we find that the most likely shape of `Oumuamua is the oblate shape with $f_l = l_1/l_3 \sim 7$. The results for other $f_l$ are shown in the embedded panel in Fig. \ref{fig:visibility}. We confirm that the probability of reproducing the observed variation amplitude \citep{Mashchenko19} peaks at $10\%$ with a 1:1:7 prolate axis ratio; at $34 \%$ with a 7:7:1 oblate axis ratio; and is $\lesssim 1.5 \%$ with a 
1:1:$5 \times 10^{-6}$ thin-solar-sail axis ratio. This indicates that the most likely shape of `Oumuamua is an oblate shape with an axis ratio of 7:7:1.  {Early studies favor a prolate shape according to the estimation of the density \citep{Meech17,Drahus18}, while \citet{Belton18} suggest an oblate shape in the case that `Oumuamua's rotational state is close to its highest energy state. By computing the possibilities of the best fitting prolate and oblate shape models, \citet{Mashchenko19} proposes that `Oumuamua is more like an oblate spheroid with some torque exerted. Here our result supports the argument that `Oumuamua is more likely to be oblate,  {making use of a different method}, which compares the possibilities of various shapes inducing the observed amplitude of the light curve.}
This low
probability of a light sail ($\sim 1.5 \%$) is due to the common occurrence of a TIS, expected
for thin  freely tumbling light sails, which leads to very low surface exposure and reflection when they are nearly 
edge-on in the direction of the  Sun (see Sect. \ref{sec:light_curve}), analogous to Saturn's ring-plane crossing. In contrast, the tumbling motion in an OIS generally does not produce
sufficiently large-amplitude modulations except under some special circumstances: $\theta + \theta^\prime \sim \pi/2$,
where $\theta$ and $\theta^\prime$ are the angles between the rotation axis relative to the normal vector of the
light sail and to the direction of the Sun, respectively.

For event B, we generated $10^4$ test light sails with initial amplitudes in the range (2.5, 3) and added random torques 
with the magnitude given by Eq. \ref{eq:taurotrad} and a random direction. The magnitude of the torque decreases 
with the distance from the Sun, and the direction is assumed constant in the corotating reference frame over time. With a 
lack of information about the torque, we made the assumption of a constant torque in the corotating frame as previous work 
obtained a good fit to the light curve by applying this assumption \citep{Mashchenko19}. Due to the torque, the light sail 
can switch from the OIS to the TIS and therefore become invisible for part of its rotation cycle. For each model, we ran a 70-day 
simulation, covering from October 25, {2017} January 2, {2018}. Since the timescale of spin evolution due to the torque is $\sim 24$ days, the rotation state of `Oumuamua since  November 15, {2017,} is considered to be random and uncorrelated to that on  {October 25, 2017,} when the light curve is derived. 
We can thus calculate the visible fraction for each case and obtain the distribution of the whole sample set, which contains $10^4$ test particles. We deduce an expectation probability of
$86.1 \pm 3.2 \%$ of the entire 49-day monitoring interval that a freely tumbling light sail's 
brightness is above the detection threshold (i.e., brighter than 27 magnitude), assuming a uniform sampling (gray distribution in Fig. \ref{fig:visibility}). If the sampling is taken at epochs when ground-based and Hubble Space Telescope (HST) astrometry was obtained, the probability would be $84.8 \pm 6.1 \%$ (blue distribution in Fig. \ref{fig:visibility}).

\begin{figure}
    \centering
    \includegraphics[width=\linewidth]{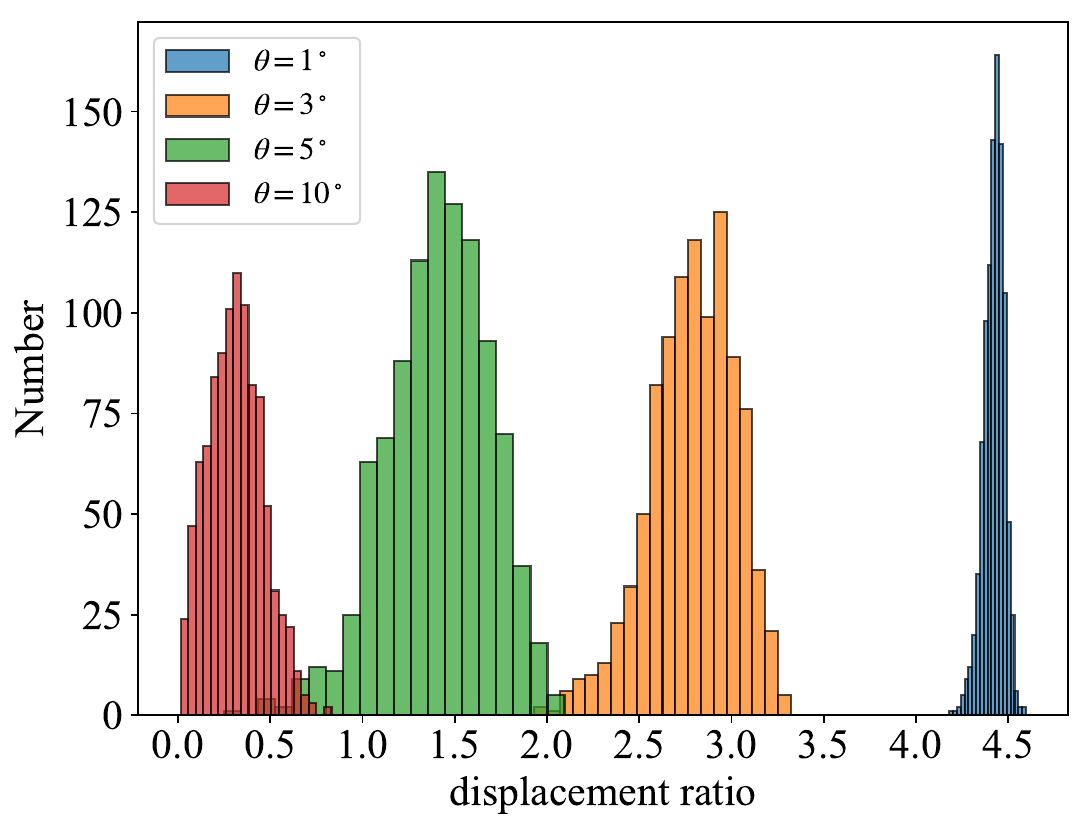}
    \caption{Distribution of the ratios of the sideways displacement to the radial displacement for different tumbling angles, $\theta$, during observation (80 days). The initial rotation states of $10^3$ test particles are constrained by the fact that the light curve amplitude is between 2.5 and 3. The sideways-to-radial displacement is $4.4 \pm 0.1$, $2.8 \pm 0.2$, $1.4 \pm 0.3$,  and $0.3 \pm 0.1$ for the initial wobble angles $\theta \leq 1^\circ $, $\leq 3^\circ$, $\leq 5^\circ$, and $\leq 10^\circ$, respectively.}

    \label{fig:fixed_theta}
\end{figure}

\begin{figure*}
    \centering
    \includegraphics[width=\linewidth]{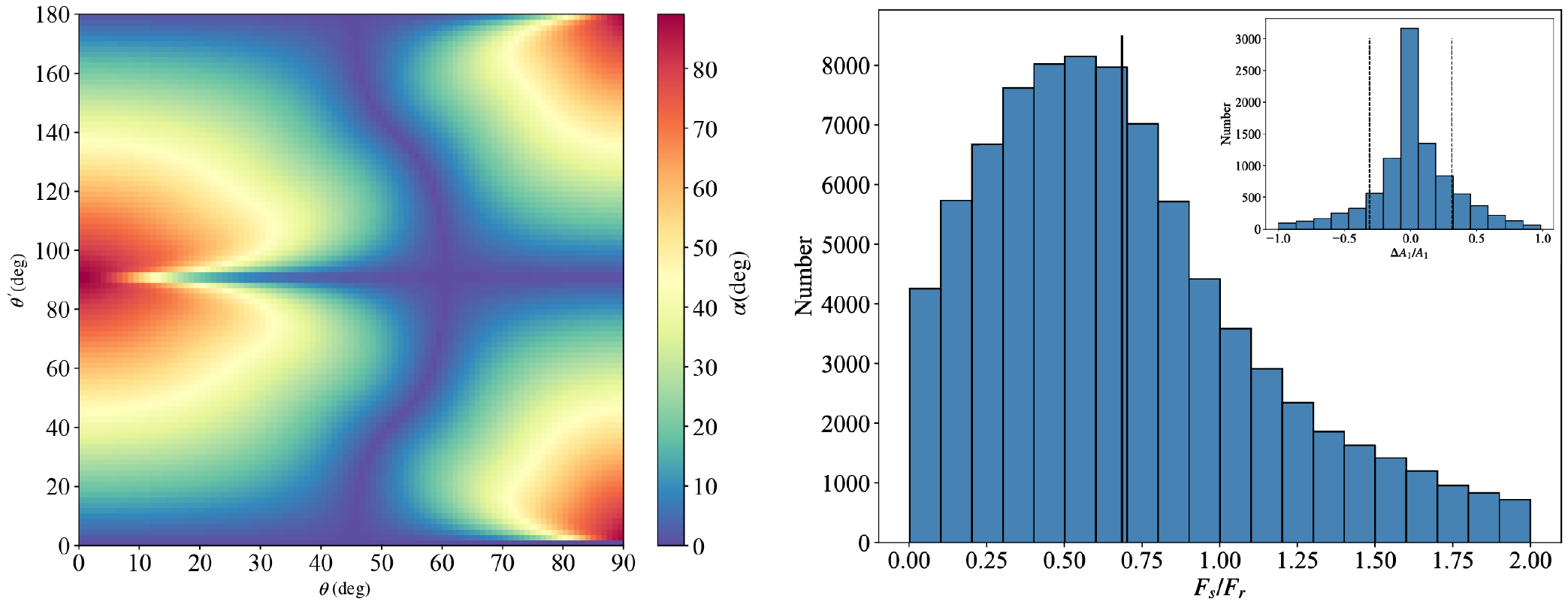}
    \caption{ {Ratio of the sideways radiation force ($F_{\rm mean,s}$) to the radial radiation force ($F_{\rm mean,r}$)} for a tumbling light sail suffering radiation pressure. (a) Distribution of the deflection angle, $\alpha$, with $\tan \alpha = {F_{\rm mean,s} /F_{\rm mean,r}}$ for the whole value range of the wobble angle, $\theta$, and the photon incident angle, $\theta^\prime$. (b) Distribution of ${F_{\rm mean,s}/ F_{\rm mean,r}}$ for $10^5$ light sails with random $\theta$ and $\theta^\prime$ after an 80-day motion.  {The mean value of the sideways-to-radial force ratio is $0.7\pm 0.4$ for light sails, while this ratio for `Oumuamua is $\sim 0.08$ \citep{Micheli18},} which is a discrepancy \citep{Micheli18} of $\sim 1.5 \sigma$. The embedded panel shows the distribution of $\Delta A_1/A_1$ for $10^4$ generated pseudo `Oumuamuas during 80-day movements,  {where $A_1$ and $\Delta A_1$  {denote} the coefficient of the radiation pressure and its variation over 55 days, from November 15, 2017, to January 2, 2018 (see Sect. \ref{sec:3}).} The standard deviation, denoted by the dash lines, is 0.3.}
    \label{fig: force_ratio}
\end{figure*}

\begin{figure*}[htbp]
\centering
\includegraphics[width = 0.75\textwidth]{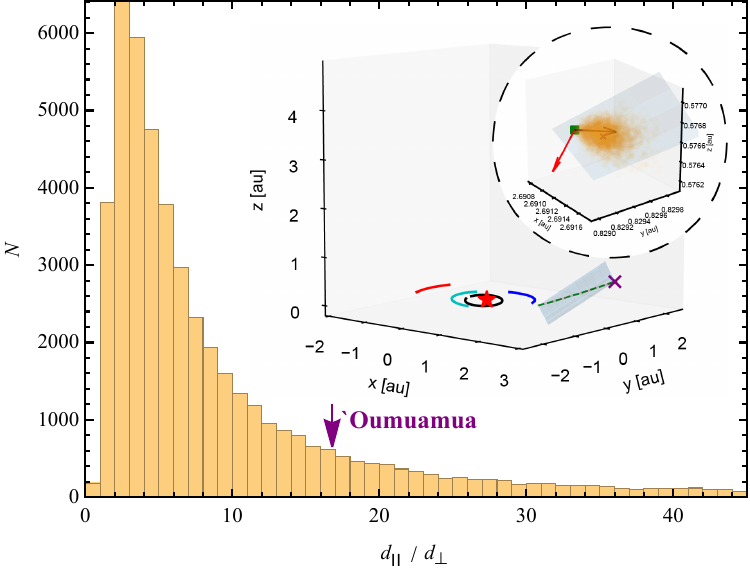}
\caption{Histogram of the ratio between the parallel deviation and the normal deviation. 
The radiation-pressure-induced deviation from a Keplerian trajectory during `Oumuamua's sojourns 
through the inner Solar System are shown with the dashed green line. The red star represents the Sun, and solid black, cyan, blue, 
and red arcs indicate the orbits of Mercury, Venus, Earth, and Mars, respectively. The purple cross indicates the position of `Oumuamua at the last record. In total, $5\times 10^4$ light-sail models with randomized initial spin orientations are integrated from `Oumuamua's discovery position for the same period of time. An additional Keplerian trajectory without radiation force is integrated, with which an orbital plane is defined. The circular inset panel zooms in to show `Oumuamua's last position (purple cross) surrounded by all light-sail models at the end of integration. And deviations from the Keplerian orbit (green square in the plane) are 
measured in the parallel and perpendicular directions to the orbital plane (indicated by two red arrows). The distribution of the ratio of the parallel displacement to the normal displacement peaks at $\sim 3,$ while `Oumuamua lies at $\sim 17$, which is obtained by applying the nongravitational acceleration suggested by \citet{Micheli18}.}
\label{fig:radiationsideways}
\end{figure*}

\begin{figure*}
    \centering
    \includegraphics[width=\linewidth]{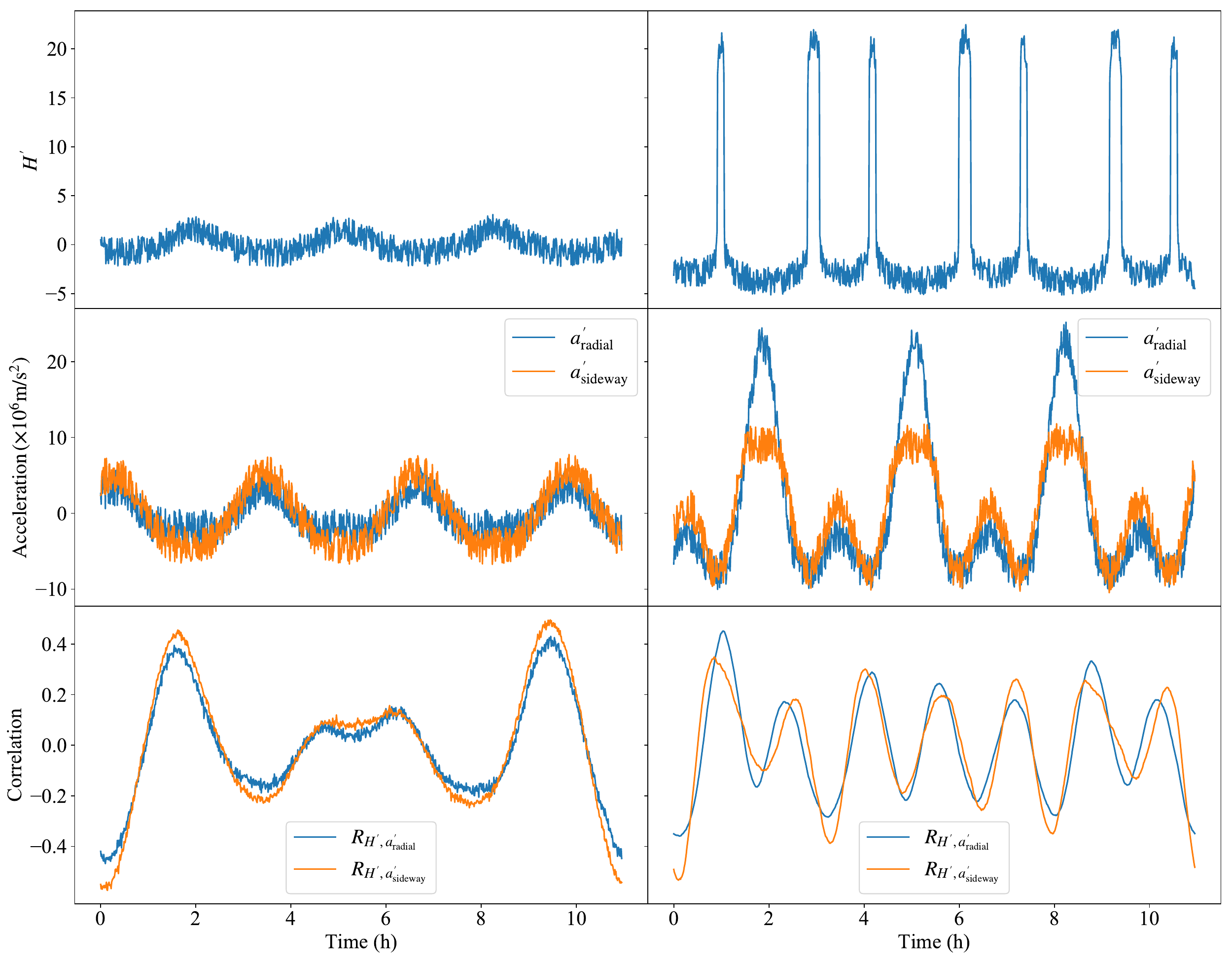}
    \caption{Light curve (upper panels), radiation acceleration (middle panels), and correlation within one day between the them (bottom panels) for the light sail in the OIS (left panels) and in the TIS (right panels). }
    \label{fig:correlation}
\end{figure*}

The likelihood of the ISO as a light sail is revealed by comparing the distribution of the visible fraction of the tested $10^4$ cases with that purely inferred from observations between  {November 15, 2017,} and  {to January 2, 2018} (49 days). The latter can be derived by statistic methods. For the simplest guess,
one could use the MLE, which yields a $\hat{\Theta}_{\rm MLE} = 55/55 = 100\%$ probability of visibility. 
However, such an estimation only generates a single fixed value. To calculate the posterior probability distribution $f(\Theta | B)$ --  that is, the probability distribution function of `Oumuamua's being visible given the fact that it was successfully detected 
in all 55 observations (event B) -- we refer to Bayesian inference for a continuous probability variable $\Theta \in [0,1]$:
\begin{equation}
    f(\Theta | B(n,k)) = \frac{f(B|\Theta) f(\Theta)}{\int_0^1 {f(B|\Theta) f(\Theta) {\rm d}\Theta}} ,
\end{equation}
where $B(n,k)$ is a statistical model describing a particular event of $k$ times positive results out of $n$ independent tests,
and here we have $n=k=55$ for event B. The $f(\Theta)$ is the prior probability distribution. Since we have no prior 
knowledge of `Oumuamua's visible probability, it is reasonable to assume $f(\Theta)=1$; in other words, any visible probability, 
$\Theta$, is equally likely. The\ $f(B|\Theta)$ is the probability of $B(n,k)$ when the probability parameter $\Theta$ is given, and it can be written as
\begin{equation}
    f(B(n,k)|\Theta) = \Theta^k (1-\Theta)^{n-k}.
\end{equation}
Thus, we can calculate the posterior probability distribution using
\begin{equation}
    f(\Theta | B(n,k)) = \frac{(n+1)!}{k!(n-k)!}\Theta^k (1-\Theta)^{n-k}.
    \label{eq:posterior}
\end{equation}
We show `Oumuamua's posterior visible probability density distribution $f(\Theta | B(55,55)$ in Fig. \ref{fig:visibility}. The expected a posteriori (EAP) of `Oumuamua's visibility is 
\begin{equation}
\begin{aligned}
\label{eq:EAP}
    \hat{\Theta}_{\rm EAP} = E(\Theta | B(55,55)) & = \int_0^1 \Theta f(\Theta | B(55,55)) {\rm d}\Theta \\
    & = \frac{56}{57} \simeq 98.2\%.
\end{aligned}
\end{equation}
This $3 \sigma$ (or $2.2 \sigma$) discrepancy between the uniform (or nonuniform) sampling model and observational detection probability poses another
challenge to the freely tumbling light-sail scenario. In fact, only 8 out of $10^4$ models report a 100\% visible fraction.

 {The above analysis could be strengthened by additional observational results, such as the {Minor Planet Center (MPC)} astrometric database, after taking into account their  limitations intrinsic to all photometric measurements obtained with  {astrometric purposes}.}


\subsection{Displacement normal to the orbit plane}
\label{sec:displacement}

`Oumuamua's ``persistently visible'' magnitude may be maintained under a highly unlikely scenario, that its 
tumbling motion is marginally kept in a protracted {OIS} near its current ($\theta+\theta^\prime \simeq \pi/2$)
configuration. This probability can be constrained by the observationally inferred nongravitational force and acceleration
in the radial ($F_{\rm r}$ away from the Sun) and sideways ($F_{\rm s}$ orthogonal to the radial) directions,
due to the reflective torque on the light sail's surface \citep{Micheli18}. Figure \ref{fig:fixed_theta} shows that for $\theta < 5^\circ-10^\circ$, $\Delta r_{\rm mean, s} >
\Delta r_{\rm mean, r}$. For a randomly distributed value of $\theta$,
the most likely mean ratio of these components $F_{\rm mean, s}/
F_{\rm mean, r} \sim 0.7 \pm 0.4$ (Eq. \ref{eq:fradsidrad}, Figs. \ref{fig: force_ratio} and 
\ref{fig:radiationsideways}), while it 
is 0.08
for the observed trajectory of `Oumuamua \citep{Micheli18}, deviating from the light-sail model by $1.5 \sigma$.  
Since these forces correlate with the light curve (Fig. \ref{fig:correlation}), which evolves 
due to changes in `Oumuamua's tumbling motion (${\dot P}_O$) and orbit, a $\sim 30 \%$ revision 
in the radiation-pressure-driven acceleration is expected during its 80 days observational window. Observation data \citep{Micheli18} show that their ratio 
is only $\sim 0.08$, while in our simulation the fraction of the ratios equal to or smaller than 0.08 is only 2\%.

As a result of the sideways force, an observable non-radial displacement is introduced to `Oumuamua's trajectory. {The sideways} force causes a displacement normal to the orbital plane. The normal displacement is a good indication of the sideways force since the radial displacement and the transverse displacement are not well defined due to their changes of direction in the orbital motion. We used the REBOUND code \citep{Rein:20120m} and the REBOUNDx extension \citep{Tamayo:2019} to compute the ratio of the parallel displacement (in-plane displacement) to the normal displacement  for $5 \times 10^4$ randomly generated models. The distribution 
of the ratio is shown in Fig. \ref{fig:radiationsideways}. The peak of the ratio distribution is at $\sim 3$, implying that the most likely parallel-to-normal ratio is $\sim 3$ for the light sail. {However, the parallel-displacement ratio of `Oumuamua was found to be $\sim 17$ by applying the nongravitational acceleration reported by \citet{Micheli18}. }
Although existing data do not have adequate accuracy to firmly establish any 
discriminating constraints on the light-sail hypothesis, they highlight the 
requirement of low-probability coincidences for `Oumuamua.

\begin{figure}
    \centering
    \includegraphics[width=\linewidth]{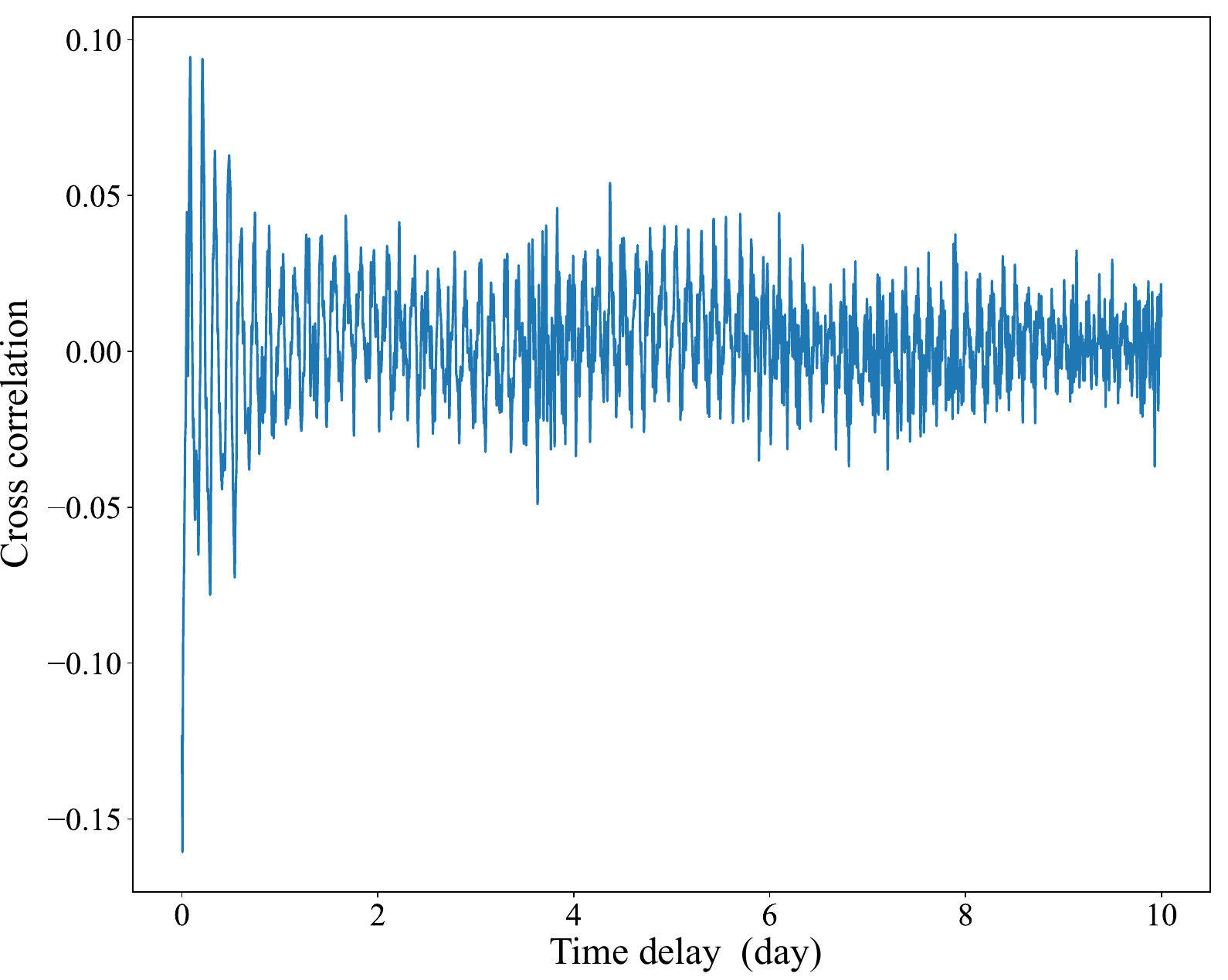}
    \caption{ {Correlation between the light curve and the acceleration as a function of the delay time for a light sail. The correlation declines quickly within two days, which indicates that high-cadence time series data are required for the diagnosis of a light sail.}}
    \label{fig:11}
\end{figure}

Since both the radiation force and the light curve highly depend on the orientation and rotation of the light sail, there is a strong correlation between the acceleration and light curve.
Figure \ref{fig:correlation} shows an example of the correlation functions between the absolute magnitude, $H'$, and the radiation force, $a'$, for the light sail. The correlation functions exhibit a regular mode, and the maximum value reaches 0.4 for both OIS and TIS, implying a high correlation between the absolute magnitude and the radiation force. However, for tumbling comets with nongravitational acceleration due to outgassing, the correlation is expected to be much weaker due to the chaotic tumbling motion and thermal delay. {This correlation for a light sail declines within several rotational cycles ($\sim 2$ days), which is shown in Fig. \ref{fig:11}; this allows us to place a requirement on the high-cadence observational data.} Therefore, with help of a huge amount of highly accurate observation data in the future, we will be able to identify the light sail by investigating the correlation between the light curve and the force curve.

\section{Summary and discussions}
\label{sec:5}
 {We have analyzed the dynamics of light sails in the interstellar space and Solar System and show that their signatures, if they exist, are observable and can be quantitatively analyzed with some forthcoming instruments, such as the Vera Rubin telescope and the Chinese Space Station Telescope (CSST).  In the forthcoming observations, our methods can function as a pipeline to test the probability of an ISO being a light sail. }

 {After applying these analyses in the context of the first-discovered ISO `Oumuamua, we find its observed properties to be  incompatible with those expected from the light-sail hypothesis. Based on inferences from `Oumuamua's data, we suggest additional stringent tests
for any potential interstellar light sails:}
\begin{description}
\item[(1)] { Freely floating light sails endure drifts due to the magnetic and gas drag by the ISM. The drift of freely rotating light sails in the ISM can reach $\sim 100\,$au over a travel distance of 1 pc. It increases with the travel distance as $\sim L^3$. 
In order for any light sail to venture into `Oumuamua's intended 
closest approach to the Sun, a constellation of light-sail probes is
needed. Some members of this flotilla should be detectable over the next 
few years.}

\item[(2)]  {The solar radiation would cause a sideways pressure if the net 
flux of incident solar light is not parallel or antiparallel to the 
light sails' orbital trajectories. Consequently, any tumbling motion 
would impose a non-negligible sideways radiation force on light sails 
and provoke sideways displacements.  In reality, the sideways force
 {experienced} by `Oumuamua as inferred from its observed orbit is one order of 
magnitude smaller than the radial force, whereas these forces are expected to be comparable for a typical tumbling light sail. Quantitatively,
the observed `Oumuamua ratio deviates from the value expected for a light 
sail by $\sim$ 1.5 $\sigma$. Moreover, the probability distribution of the 
ratio between the displacement in the orbital plane to that in the 
normal direction for light sails peaks at 3, while the 
observationally inferred ratio is 17 for `Oumuamua.
The displacement normal to the orbital plane is expected 
to be correlated with the light 
curve of light sails. This test can be applied with future 
high-cadence observations.}

\item[(3)]  {The amplitude of the light curve can also be used to
assess the likelihood of the axis ratio of ISOs. Our analysis of `Oumuamua's observed light curve suggests it is most likely to 
have an oblate shape with an axis ratio of $\sim 7:7:1,$  {while the probability for such a light curve to be due to a light sail} is $\sim 1.5\%$. The large 
amplitudes in the light-curve variation and the prospect of 
undetectable phases during their tumbling motion can be used to
quantify the light-sail probability for future observations of
ISOs.  The probability for a light sail to be 
persistently brighter than the 27th magnitude in all 55 `Oumuamua's 
monitoring attempts during the two 
months after its discovery is estimated to be 0.4\%. We find
a $3\sigma$ discrepancy between `Oumuamua's perfect-detection record 
and the expectation based on the light-sail hypothesis.\\ }

\end{description}

We suggest that `Oumuamua is unlikely to be a light sail. The dynamics of an intruding light sail would give rise to observational features such as the occurrence of its cohorts, a remarkable sideways radiation force and normal displacement from the Kepler orbit, a light curve with an extremely high amplitude, and irregular invisibility in the sky. These features can be quantitatively identified and analyzed with our methods in future surveys.

\begin{acknowledgements}
   We thank Greg Laughlin and Scott Tremaine, as well as Tingtao Zhou, Patrick Michel, Meng Su, Xiaojia Zhang, Wenchao Wang, 
   Yue Wang, Quanzhi Ye, Shoucun Hu, and Bo Ma for useful discussions. This work is supported by the National Natural Science Foundation of China under grant No. 11903089, the Guangdong Basic and Applied Basic Research Foundation under grant Nos. 2021B1515020090 and 2019B030302001, the Fundamental Research Funds for the Central Universities, Sun Yat-sen University under grant No. 22lgqb33, and the China Manned Space Project under grant Nos. CMS-CSST-2021-A11 and CMS-CSST-2021-B09.
   \end{acknowledgements}

%
%
\bibliographystyle{aa} 
\bibliography{44119corr}

\begin{thebibliography}{45}
\expandafter\ifx\csname natexlab\endcsname\relax\def\natexlab#1{#1}\fi

\bibitem[{Abell {et~al.}(2009)Abell, Allison, Anderson, Andrew, Angel, Armus,
  Arnett, Asztalos, Axelrod, Bailey, {et~al.}}]{Abell2009}
Abell, P.~A., Allison, J., Anderson, S.~F., {et~al.} 2009, arXiv preprint
  arXiv:0912.0201

\bibitem[{Bannister {et~al.}(2019)Bannister, Bhandare, Dybczy{\'n}ski,
  Fitzsimmons, Guilbert-Lepoutre, Jedicke, Knight, Meech, McNeill, Pfalzner,
  {et~al.}}]{Bannister19}
Bannister, M.~T., Bhandare, A., Dybczy{\'n}ski, P.~A., {et~al.} 2019, Nature
  astronomy, 3, 594

\bibitem[{Bannister {et~al.}(2017)Bannister, Schwamb, Fraser, Marsset,
  Fitzsimmons, Benecchi, Lacerda, Pike, Kavelaars, Smith,
  {et~al.}}]{Bannister17}
Bannister, M.~T., Schwamb, M.~E., Fraser, W.~C., {et~al.} 2017, The
  Astrophysical Journal Letters, 851, L38

\bibitem[{Belton {et~al.}(2018)Belton, Hainaut, Meech, Mueller, Kleyna, Weaver,
  Buie, Drahus, Guzik, Wainscoat, {et~al.}}]{Belton18}
Belton, M.~J., Hainaut, O.~R., Meech, K.~J., {et~al.} 2018, The Astrophysical
  Journal Letters, 856, L21

\bibitem[{Bialy \& Loeb(2018)}]{BialyLoeb18}
Bialy, S. \& Loeb, A. 2018, The Astrophysical Journal, 868, L1

\bibitem[{Bolin {et~al.}(2017)Bolin, Weaver, Fernandez, Lisse, Huppenkothen,
  Jones, Juri{\'c}, Moeyens, Schambeau, Slater, {et~al.}}]{Bolin17}
Bolin, B.~T., Weaver, H.~A., Fernandez, Y.~R., {et~al.} 2017, The Astrophysical
  Journal Letters, 852, L2

\bibitem[{Bottke~Jr {et~al.}(2006)Bottke~Jr, Vokrouhlick{\`y}, Rubincam, \&
  Nesvorn{\`y}}]{Bottke06}
Bottke~Jr, W.~F., Vokrouhlick{\`y}, D., Rubincam, D.~P., \& Nesvorn{\`y}, D.
  2006, Annu. Rev. Earth Planet. Sci., 34, 157

\bibitem[{Choudhuri \& Roy(2019)}]{Choudhuri19}
Choudhuri, S. \& Roy, N. 2019, Monthly Notices of the Royal Astronomical
  Society, 483, 3437

\bibitem[{Crutcher(2012)}]{Crutcher2012}
Crutcher, R.~M. 2012, Annual Review of Astronomy and Astrophysics, 50, 2012

\bibitem[{{\'C}uk(2018)}]{Cuk18}
{\'C}uk, M. 2018, The Astrophysical Journal Letters, 852, L15

\bibitem[{Curran(2021)}]{Curran2021}
Curran, S. 2021, arXiv preprint arXiv:2105.09435

\bibitem[{Do {et~al.}(2018)Do, Tucker, \& Tonry}]{Do18}
Do, A., Tucker, M.~A., \& Tonry, J. 2018, The Astrophysical Journal Letters,
  855, L10

\bibitem[{Drahus {et~al.}(2018)Drahus, Guzik, Waniak, Handzlik, Kurowski, \&
  Xu}]{Drahus18}
Drahus, M., Guzik, P., Waniak, W., {et~al.} 2018, Nature Astronomy, 2, 407

\bibitem[{Draine(2010)}]{Draine10}
Draine, B.~T. 2010, Physics of the interstellar and intergalactic medium,
  Vol.~19 (Princeton University Press)

\bibitem[{{Draine}(2011)}]{Draine11}
{Draine}, B.~T. 2011, {Physics of the Interstellar and Intergalactic Medium}
  (Princeton University Press)

\bibitem[{Drell {et~al.}(1965)Drell, Foley, \& Ruderman}]{Drell65}
Drell, S.~D., Foley, H.~M., \& Ruderman, M.~A. 1965, Journal of Geophysical
  Research (1896-1977), 70, 3131

\bibitem[{Flekk{\o}y {et~al.}(2019)Flekk{\o}y, Luu, \& Toussaint}]{Flekkoy19}
Flekk{\o}y, E.~G., Luu, J., \& Toussaint, R. 2019, The Astrophysical Journal
  Letters, 885, L41

\bibitem[{Fraser {et~al.}(2018)Fraser, Pravec, Fitzsimmons, Lacerda, Bannister,
  Snodgrass, \& Smoli{\'c}}]{Fraser18}
Fraser, W.~C., Pravec, P., Fitzsimmons, A., {et~al.} 2018, Nature Astronomy, 2,
  383

\bibitem[{Hsieh {et~al.}(2021)Hsieh, Laughlin, \& Arce}]{Hsieh21}
Hsieh, C.-H., Laughlin, G., \& Arce, H.~G. 2021, The Astrophysical Journal,
  917, 20

\bibitem[{Jackson \& Desch(2021)}]{Jackson21}
Jackson, A.~P. \& Desch, S.~J. 2021, Journal of Geophysical Research: Planets,
  126, e2020JE006706

\bibitem[{Jewitt {et~al.}(2017)Jewitt, Luu, Rajagopal, Kotulla, Ridgway, Liu,
  \& Augusteijn}]{Jewitt17}
Jewitt, D., Luu, J., Rajagopal, J., {et~al.} 2017, The Astrophysical Journal
  Letters, 850, L36

\bibitem[{Katz(2019)}]{Katz2019}
Katz, J. 2019, Astrophysics and Space Science, 364, 1

\bibitem[{Katz(2021)}]{Katz2021}
Katz, J. 2021, arXiv preprint arXiv:2102.07871

\bibitem[{Knight {et~al.}(2017)Knight, Protopapa, Kelley, Farnham, Bauer,
  Bodewits, Feaga, \& Sunshine}]{Knight17}
Knight, M.~M., Protopapa, S., Kelley, M.~S., {et~al.} 2017, The Astrophysical
  Journal Letters, 851, L31

\bibitem[{Laibe \& Price(2012)}]{Laibe12}
Laibe, G. \& Price, D.~J. 2012, Monthly Notices of the Royal Astronomical
  Society, 420, 2345

\bibitem[{Loeb(2021)}]{Loeb21}
Loeb, A. 2021, Extraterrestrial: The First Sign of Intelligent Life Beyond
  Earth (Houghton Mifflin)

\bibitem[{Mashchenko(2019)}]{Mashchenko19}
Mashchenko, S. 2019, Monthly Notices of the Royal Astronomical Society, 489,
  3003

\bibitem[{Meech {et~al.}(2017)Meech, Weryk, Micheli, Kleyna, Hainaut, Jedicke,
  Wainscoat, Chambers, Keane, Petric, {et~al.}}]{Meech17}
Meech, K.~J., Weryk, R., Micheli, M., {et~al.} 2017, Nature, 552, 378

\bibitem[{Micheli {et~al.}(2018)Micheli, Farnocchia, Meech, Buie, Hainaut,
  Prialnik, Sch{\"o}rghofer, Weaver, Chodas, Kleyna, {et~al.}}]{Micheli18}
Micheli, M., Farnocchia, D., Meech, K.~J., {et~al.} 2018, Nature, 559, 223

\bibitem[{{Moro-Mart{\'\i}n}(2019)}]{Moro-Martin2019}
{Moro-Mart{\'\i}n}, A. 2019, The Astrophysical Journal Letters, 872, L32

\bibitem[{Muinonen \& Lumme(2015)}]{Muinonen15}
Muinonen, K. \& Lumme, K. 2015, Astronomy \& Astrophysics, 584, A23

\bibitem[{Pfalzner {et~al.}(2020)Pfalzner, Davies, Kokaia, \&
  Bannister}]{Pfalzner20}
Pfalzner, S., Davies, M.~B., Kokaia, G., \& Bannister, M.~T. 2020, The
  Astrophysical Journal, 903, 114

\bibitem[{Portegies~Zwart {et~al.}(2018)Portegies~Zwart, Torres, Pelupessy,
  Bédorf, \& Cai}]{Portegies18}
Portegies~Zwart, S., Torres, S., Pelupessy, I., Bédorf, J., \& Cai, M.~X.
  2018, Monthly Notices of the Royal Astronomical Society: Letters, 479, L17

\bibitem[{Raymond {et~al.}(2018{\natexlab{a}})Raymond, Armitage, \&
  Veras}]{Raymond18b}
Raymond, S.~N., Armitage, P.~J., \& Veras, D. 2018{\natexlab{a}}, The
  Astrophysical Journal Letters, 856, L7

\bibitem[{Raymond {et~al.}(2018{\natexlab{b}})Raymond, Armitage, Veras,
  Quintana, \& Barclay}]{Raymond18a}
Raymond, S.~N., Armitage, P.~J., Veras, D., Quintana, E.~V., \& Barclay, T.
  2018{\natexlab{b}}, Monthly Notices of the Royal Astronomical Society, 476,
  3031

\bibitem[{Rein \& Liu(2012)}]{Rein:20120m}
Rein, H. \& Liu, S.-F. 2012, Astronomy and Astrophysics, 537, 128

\bibitem[{Rubincam(2000)}]{Rubincam00}
Rubincam, D.~P. 2000, Icarus, 148, 2

\bibitem[{Stern(1986)}]{Stern86}
Stern, S.~A. 1986, Icarus, 68, 276

\bibitem[{Tamayo {et~al.}(2019)Tamayo, Rein, Shi, \& Hernandez}]{Tamayo:2019}
Tamayo, D., Rein, H., Shi, P., \& Hernandez, D.~M. 2019, Monthly Notices of the
  Royal Astronomical Society, 491, 2885

\bibitem[{Trilling {et~al.}(2018)Trilling, Mommert, Hora, Farnocchia, Chodas,
  Giorgini, Smith, Carey, Lisse, Werner, {et~al.}}]{Trilling18}
Trilling, D.~E., Mommert, M., Hora, J.~L., {et~al.} 2018, The Astronomical
  Journal, 156, 261

\bibitem[{Trilling {et~al.}(2017)Trilling, Robinson, Roegge, Chandler, Smith,
  Loeffler, Trujillo, Navarro-Meza, \& Glaspie}]{Trilling17}
Trilling, D.~E., Robinson, T., Roegge, A., {et~al.} 2017, The Astrophysical
  Journal Letters, 850, L38

\bibitem[{Tsuda {et~al.}(2011)Tsuda, Mori, Funase, Sawada, Yamamoto, Saiki,
  Endo, \& Kawaguchi}]{Tsuda11}
Tsuda, Y., Mori, O., Funase, R., {et~al.} 2011, Acta astronautica, 69, 833

\bibitem[{Vokrouhlick{\`y} \& {\v{C}}apek(2002)}]{Vokrouhlicky02}
Vokrouhlick{\`y}, D. \& {\v{C}}apek, D. 2002, Icarus, 159, 449

\bibitem[{Zhang \& Lin(2020)}]{Zhang20}
Zhang, Y. \& Lin, D.~N. 2020, Nature Astronomy, 4, 852

\bibitem[{Zhou(2020)}]{Zhou20}
Zhou, W.~H. 2020, The Astrophysical Journal, 899, 42

\end{thebibliography}






   
  




\appendix

\section{ISM's torque on the light sail's asymmetric surface and `Oumuamua's spin}
\label{sec:ismtorque}
 
The large-amplitude variations in `Oumuamua's light curve suggest it has an asymmetry surface. 
Such a feature may be inherent in its original architecture, be bestowed by micro-meteorite 
and cosmic-ray bombardments during its passage through the ISM, or be deformed by the thermal 
stress due to a nonuniform albedo.  Although the light sail may also have encountered an air 
mass comparable to $m_{\rm O}$ after traveling $L \gtrsim L_{\rm term, gas}$ \citep{BialyLoeb18}, 
the ISM's condensation on its surface is unlikely to significantly modify its thickness, $l_3$, 
especially in the limit $L \lesssim L_{\rm term, gas}$.

\begin{figure*}
    \centering
    \includegraphics[width=\linewidth]{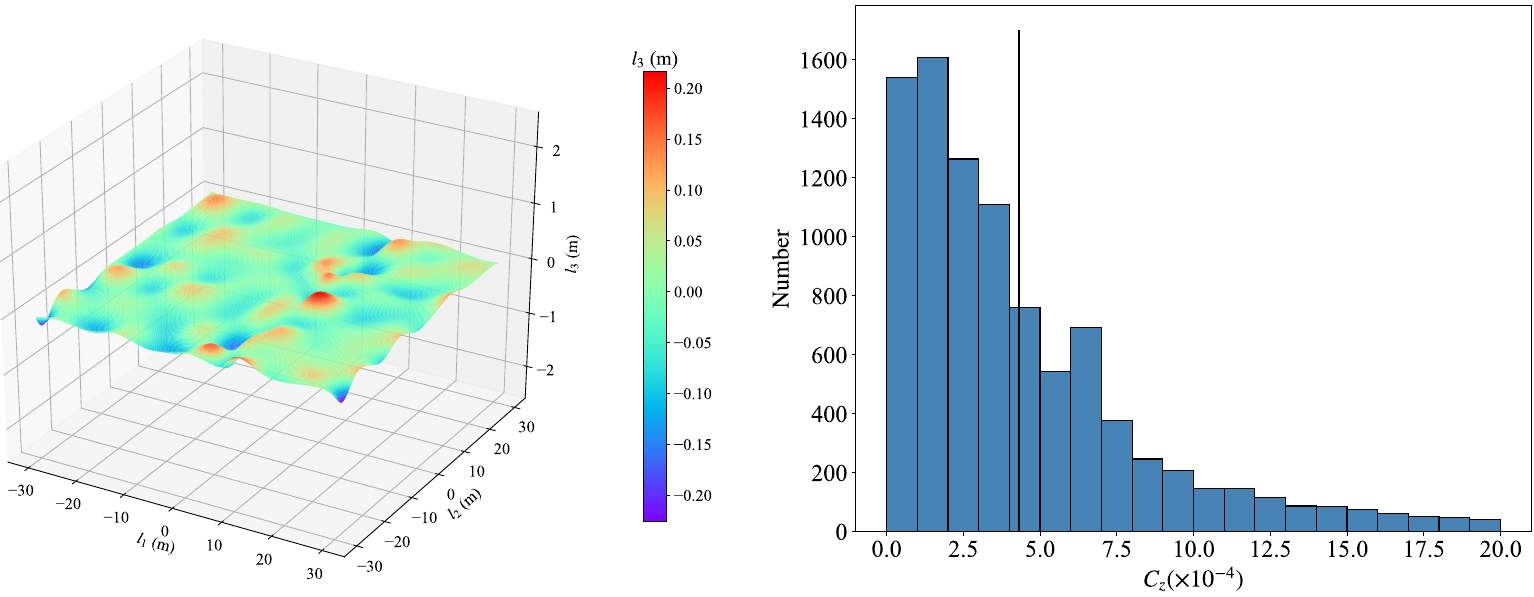}
    \caption{Example of the rough surface with $\Delta h \sim 0.05 ~\rm m$ (left panel) and the distribution of the values of $C_z$ for $10^4$ runs on random surfaces with $\Delta h \approx 0.05 \rm ~m$. The mean value of $C_z$ is $(4 \pm 4) \times 10^{-4}$.}
    \label{fig:vertex}
\end{figure*}

With an asymmetric shape, the light sail endures a torque due to the gas ram
pressure and dust bombardment. Over time, this torque changes the light sail's 
spin orientation and frequency \citep{Zhou20}. We approximate the torque due to gas 
pressure by
\begin{equation}
    T_{\rm gas} \simeq C_{z,\rm{ISM}} m_p n_{\rm ISM} v^2 l_1^3
\label{eq:torquegas}
,\end{equation}
where $C_{z,\rm{ISM}}$ measures the efficiency of the gas torque, depending on 
light sail's shape and surface topology. Similarly, torque due to dust bombardment \citep{Stern86} is
\begin{equation}
    T_{\rm dust} \simeq C_{z,\rm{ISM}} \bar \mu_{\rm gr} (h_1 + h_2) v v_{\rm ej} l_1^3
,\end{equation}
where $\bar \mu_{\rm gr}$ is the mean mass density of the interstellar grains, 
$h_1$ and $h_2$ are the ratios of evaporated mass and the ejected mass of the object 
to the mass of incident grain, respectively, and $v_{\rm ej}$ is the velocity of the ejection. The dust torque, $T_{\rm dust}$, does not make much difference unless `Oumuamua moves very fast (e.g., $> 20 ~\rm km/s$).

In order to infer the magnitude of $C_{z,\rm{ISM}}$, we note that the irregular roughness 
and/or curvatures on the light sail's surface also lead to a finite radiative torque, and the torque coefficient $C_{z,\rm{ISM}} \simeq C_{z,\rm{rad}}$ \citep{Zhou20}. 
The finite magnitude of the radiative torque arises 
from a residual radiation pressure exerted by 
the incident photons on an asymmetric surface that does not cancel out during any spin 
period \citep{Rubincam00, Vokrouhlicky02, Bottke06}. Averaged over time, the radiative torque is approximately 
\begin{equation}   
T_{\rm rad} \sim {C_{z,\rm{rad}} \Phi A l_1 \over c }
\label{eq:radtorque}
,\end{equation}
where the efficiency of the radiative torque $C_{z,\rm{rad}} \simeq C_{z,\rm{ISM}}$,
$c$ is the light speed, $A = f_{\rm A} l_1 l_2$, and $\Phi=L_\odot/4 \pi r^2$ is the solar flux at 
the object's heliocentric distance, $r$.
Subject to this torque, the light sails spin to evolve on a timescale 
\begin{equation}  
\tau_{\rm rad} \simeq {\omega I_3 \over T_{\rm rad} } \simeq {\omega \rho_{\rm O} l_1 l_3  c \over 8 C_{z,\rm{rad}} \Phi f_{\rm A}}
\label{eq:taurotrad}
,\end{equation}
where its moment of inertia $I_3 \sim \rho_{\rm O} l_1^3 l_2 l_3$.

`Oumuamua's observed de-spin timescale ($P_{\rm O}/{\dot P}_{\rm O} \sim 24~$days) during its passage through the inner Solar System
has been explained in terms of this radiative torque \citep{Mashchenko19}. Applying our model parameters to Eq.
(\ref{eq:taurotrad}), we infer $C_{z,\rm{rad}}$ (as well as $C_{z,\rm{ISM}}) \sim 4 \times 10^{-4}$. 
From this value of both $C_{z, \rm {rad}}$ and $C_{z,\rm{ISM}}$, we place constraints on the surface roughness of
the light sail with a set of  Monte Carlo simulations.  For our numerical model of the light sail shape, we 
constructed hundreds of surfaces with $\sim 1500$ vertexes and $\sim 3000$ facets in each 
generated surface (as shown in Fig. \ref{fig:solarsail_rotation}). The roughness can be measured by the standard deviation of 
departures, $\Delta h$, from vertexes to a hypothetical perfectly flat sheet. We tested different cases where 
the deviations are $\Delta h \simeq (0.1, 1, 10, 10^2) l_3 $. 
For each case, we processed 10000 runs with random surfaces and random orientations. 
Figure \ref{fig:vertex} shows a  typical example of the distribution 
of the departures of vertexes from the plane with the standard deviation of 50 mm. The results show that the surface cannot introduce a sufficiently large torque efficiency ($C_{z, {\rm ISM}} = 4 \times 10^{-4}$) 
unless it has a roughness such that the deviation of the vertexes is hundreds of times greater than the 
light sail thickness ($l_3\sim 3-5 \times 10^{-4} {\rm m}$). The results are shown in Fig. \ref{fig:vertex}.

\section{{Spin evolution during passage through the ISM}}
\label{sec:spinism}

During its passage through the ISM with $v\lesssim 20$ km s$^{-1}$, the light sail's spin 
evolution is mostly determined by the gas torque \citep{Zhou20}, $T_{\rm gas}$  (Eq. \ref{eq:torquegas}).  
The maximum moment of inertia, $I \approx m l_1^2/8$, of a solar sail (with a large dimension ratio 
$l_1/l_3 > > 1$) dominates over that of its command module with sizes $\lesssim 0.1 l_1$. (The actual 
integrated moment of inertia of the spacecraft is likely to be slightly smaller than this value since 
part of the mass is contained in the command module located near the center of the spacecraft.) 
The angular acceleration of the space craft is 
\begin{equation}
    \dot \omega \simeq {T_{\rm gas} \over I} = { 8 C_{z,\rm{ISM}} m_p n_{\rm ISM} v^2  \over \rho_{\rm O} l_1^2  }.
\end{equation}
The light sail's spin angular momentum and its direction of acceleration, $a_{\rm gas}$, change on the timescale of 
\begin{equation}
    \tau_{\rm{rot}} \simeq {\omega \over \dot \omega} 
    \sim {\omega \rho_{\rm O} l_1l_3 \over 
    8 C_{z,\rm{ISM}} m_p n_{\rm ISM} v^2} \sim {\omega l_1 \tau_{\rm{term,gas}} \over 
    8 C_{z,\rm{ISM}} v} .
    \label{eq:taurot}
\end{equation}
Since the coefficient of the gas torque, $C_{z,\rm{ISM}}$, has the same order of magnitude of that of the radiative 
torque, $C_{z,\rm{rad}}$ \citep{Zhou20}, we estimate $C_{z,\rm{ISM}} \simeq C_{z,\rm{rad}} 
\simeq 4 \times 10^{-4}$ from `Oumuamua's observed de-spin rate during its passage through 
the Solar System (Appendix A). 

It is possible for the $T_{\rm gas}$ torque to spin down the light sail, deplete its spin angular momentum, and restart 
a subsequent rotational cycle with a different orientation. It is also possible for the torque to transfer positive 
angular momentum to the object until it reaches a breakup spin limit. A creative epilogue \citep{BialyLoeb18, Loeb21}
of the spacecraft hypothesis is the suggestion that `Oumuamua's tumbling motion may have been intentionally instated  
to enable the spacecraft to emit and receive signals in and from all directions. Since any initial spin state
can only be preserved over $\tau_{\rm{rot}} \sim 1 {( \rho_{\rm O} l_3 / 1 {\rm kg \ m^{-2}} )} {(10^6  {\rm m^{-3}} / n_g )} 
{( l_1 / 60 {\rm m} )} \rm Myr$, such an assumption would require that `Oumuamua's interstellar odyssey, with its observed $v_{\rm O}$,
be limited to a maximum traveled distance $L_{\rm max} \approx 10 \, \rm pc$. The total population of stars, including white
dwarfs and brown dwarfs, in this domain is $\sim 400$.

\section{Accumulative drift}
\label{sec:accumulative}
\begin{figure}
    \centering
    \includegraphics[width=\linewidth]{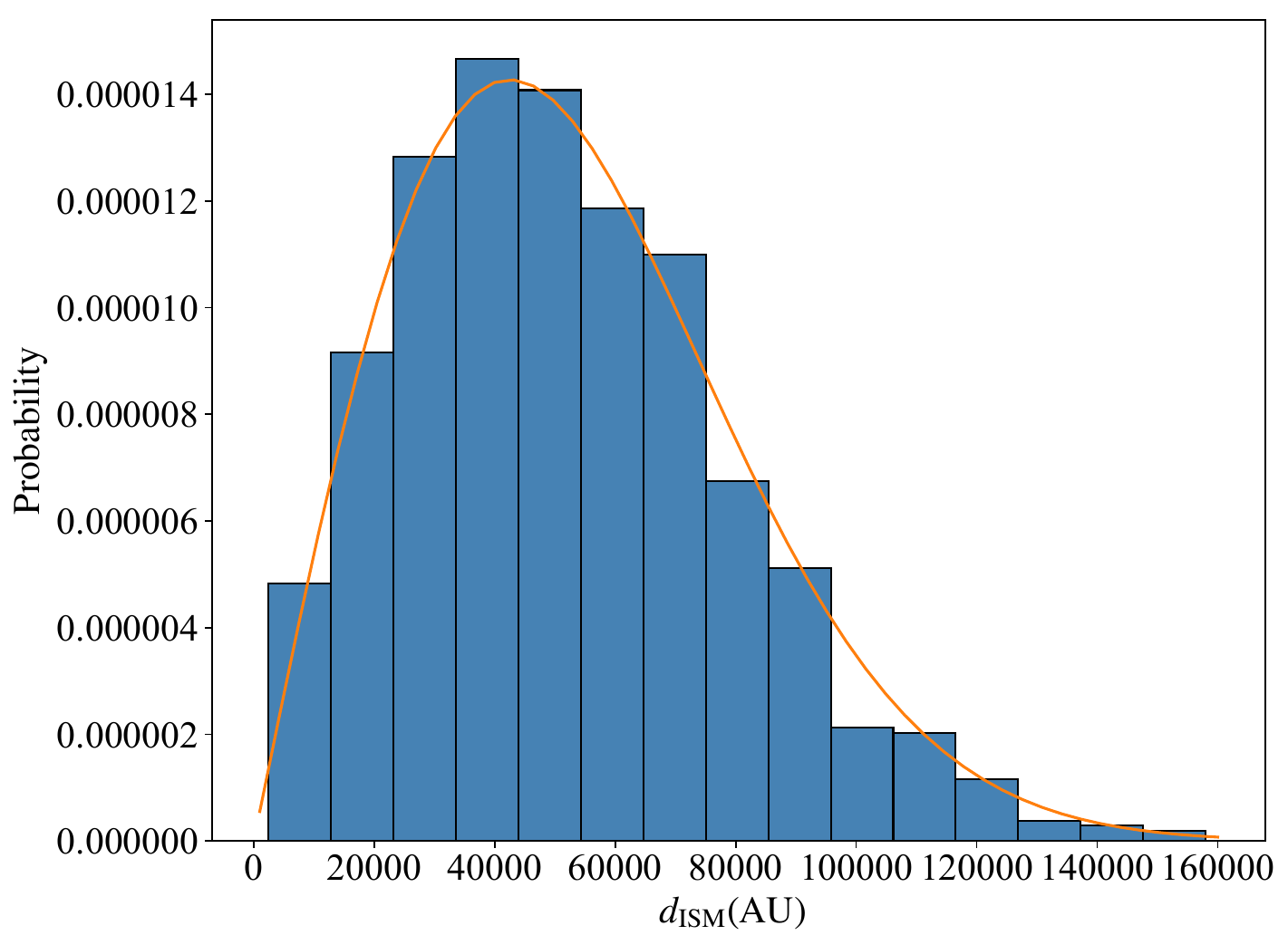}
    \caption{Normalized distribution of $d_{\rm ISM}$ for 1000 test particles that feel a random magnetic force when the travel distance is $100$ pc. The orange curve is the Rayleigh distribution given by Eq. \ref{eq:rayleigh}.}
    \label{fig:random_walk}
\end{figure}

In the limit $t \geq t_{\rm rot}$ and $t \geq t_{\rm tur}$, 
${\rm E}(d_{\rm gas})  $ and ${\rm E}(d_{\rm mag})  $  (in Eqs. \ref{eq:dgas} and \ref{eq:dmag}) 
are derived for randomly evolving acceleration $a_{\rm gas}$ and $a_{\rm mag}$. 
For the convenience of a general discussion, we use variables $\{{\rm E}(d_{\rm rand}), 
\Delta t, a_{\rm rand} \}$ to represent both $\{{\rm E}(d_{\rm gas})  , \tau_{\rm rot}, 
a_{\rm drfit} \}$ and $\{{\rm E}(d_{\rm mag})  , \tau_{\rm tur}, a_{\rm mag} \}$. 
The motion of the object follows Langevin's equation:\begin{equation}
        \vec {\ddot {d}} = -\lambda \vec v + \vec a_{\rm rand},\end{equation}
where the drag coefficient $\lambda = 0$ in our case. We started with the simplest 1D model, where the acceleration $a_{{\rm rand},i}$ is reset, after each time step $\Delta t$, to be $\pm a_{\rm rand}$ with a 50\% probability 
for each value. The velocity gains a random increment $a_{{\rm rand},i} \Delta t$ in each time interval $\Delta t$. The velocity at time $t$ is the sum of the independent random variable $a_{{\rm rand},i}$ for $i = 0,1,...N_{\rm rand}$ with $N_{\rm rand} = t/\Delta t$. According to the central limit theorem, the probability distribution of velocity, $v_{{\rm rand}}$, obeys a Gaussian distribution: $\mathcal N(0,  N_{\rm rand} a_{\rm rand}^2 \Delta t^2)$. The displacement is 
\begin{equation}
d_{\rm rand} = \int_0^{\tau} v d\tau.
\end{equation}
The distribution of $d_{\rm rand}$ can be solved by determining the distribution of the limit of the right-hand sums:
\begin{equation}
\label{eq: d_rand}
d_{\rm rand} =  \sum_{i = 0}^{N_{\rm rand} -1 } (v_i \Delta t )  \ \ \ \  
    {\rm where} \  \ \  \ v_i = \sum_{k = 0}^{i - 1 } a_{\rm rand,k} \Delta t.
\end{equation}
Since only the acceleration is the independent variable, we needed to unfold Eq \ref{eq: d_rand} in terms of $a_{{\rm rand},i}$:
\begin{equation}
    \begin{aligned}
    d_{\rm rand}  &= \Delta t(v_{\rm rand, 0} + v_{\rm rand,1} + ... +v_{{N_{\rm rand} -1}}) \\
& = \Delta t^2[a_{\rm rand,0}+(a_{\rm rand,0}+a_{\rm rand,1}) +...  \\
& ~~~~+(a_{\rm rand,0}+a_{\rm rand,1}+...+a_{{N_{\rm rand} -2}})] \\
& = \Delta t^2 (N_{\rm rand}-1)a_{\rm rand,0} + (N_{\rm rand}-2)a_{\rm rand,1} +...+ a_{{N_{\rm rand} -2}} \\
& = \Delta t^2 \sum_{i=0}^{N_{\rm rand} -2} (N_{\rm rand} -1-i)a_{i} 
    \end{aligned}
.\end{equation}
Due to the central limit theorem, the sum of independent random variables $a_{{\rm rand},i}$ (with the variance of $a_{\rm rand}$) exhibits a Gaussian distribution, with the variance
\begin{equation}
    \begin{aligned}
{\rm Var}(d_{\rm rand})& = {\rm Var}( \sum_{i=0}^{N_{\rm rand} -2} (N_{\rm rand} -1-i)a_{{\rm rand},i} \Delta t^2 ) \\
& =  \sum_{i=0}^{N_{\rm rand} -2} [(N_{\rm rand} -1-i)a_{\rm rand} \Delta t^2 )]^2 \\
& =a_{\rm rand}^2 \Delta t^4 [1+2^2+...+(N_{\rm rand} -1)^2]\\
& \simeq {1\over 3} a_{\rm rand}^2 \Delta t^4 N_{\rm rand}^3
    \end{aligned}
.\end{equation}
Therefore, the displacement $d_{\rm rand} \sim \mathcal N(0,  a_{\rm rand}^2 \Delta t^4 N_{\rm rand}^3/3)$. The expected value ${\rm E}(\vert d_{\rm rand} \vert) \sim  a_{\rm rand} \Delta t^2 N_{\rm rand}^{3/2}/\sqrt{3}$.
The displacement in higher dimensions obeys a chi distribution. In particular in a 2D case, the probability distribution of the 
displacement,  $d$, is given by a Rayleigh distribution:  
\begin{equation}
    p(d) = {2d \over N_{\rm rand}^3 \Delta d^2} \exp\left(-{d^2 \over N_{\rm rand}^3 \Delta d^2 }  \right),
        \label{eq:rayleigh}
\end{equation} 
with $\Delta d = a_{\rm rand} \Delta t^2 /\sqrt{3}$. This analytic approximation of $p(d)$ was verified with a Monte Carlo simulation for the random-walk process 
(Fig. \ref{fig:random_walk}). The expected value is expressed as
\begin{equation}
{\rm E}(d_{\rm rand}) = a_{\rm rand} \Delta t^2 N_{\rm rand}^{3/2} /\sqrt{6} = a_{\rm rand} \Delta t^{1/2} t^{3/2}/\sqrt{6}.
\label{eq:erand}
\end{equation}

\section{Radiation force for a tumbling light sail}

\subsection{Coordinate system}
In order to reconstruct `Oumuamua's equation of motion in the Solar System, 
we approximated it as an object with a non-principal axis rotation analogous to that of 
a symmetric top. In this case, the spin axis rotates around the angular momentum vector,
${\vec J}$. At a heliocentric position, ${\vec r}$,  the force components along the direction 
$\vec r \times \vec J$ cancel each other out.
The resulting net force by radiation pressure lies on the ($\vec r$, $\vec J$) plane. 
The mean force can be considered in a coordinate system described by three basis vectors, $\vec e_X$, 
$\vec e_Y$, and $\vec e_Z$, with 
\begin{equation}
    \left \{
    \begin{aligned}
        & \vec e_Z = {\vec J \over J} \\
        & \vec e_Y = {\vec e_Z \times \vec r \over r}\\
        & \vec e_X = {\vec e_Y \times \vec e_Z}.
    \end{aligned}
        \right .
\end{equation}
Since `Oumuamua's spin frequency, $\omega_{\rm O}$, is much faster than its heliocentric orbital angular frequency, 
we can approximate $\vec e_r$ to be a constant within a rotational period. We also need a rotating 
coordinate frame, described by the three unit vectors $\vec e_x$, $\vec e_y$, and $\vec e_z$ along the three 
principal axes of `Oumuamua. Here $\vec e_z$ points in the positive direction of the principal axis 
of the maximum moment of inertia. The orientation of the rotating coordinate frame can be described 
by three Euler angles, $\phi$, $\theta,$ and $\psi$ (in $x-z-x$ sequence). For an arbitrary vector 
$\vec k$, the transformation between the nonrotating reference frame ($X Y Z$) and the rotating 
reference frame ($x y z$) is
$    \vec k = R(\phi,\theta,\psi) \vec k^\prime.$
Here $\vec k $ denotes the vector in the nonrotating reference frame while $\vec k^\prime$ denotes the vector 
in the rotating frame. Here $R(\phi,\theta,\psi)$ is the rotation matrix.

In the spaceship hypothesis, `Oumuamua is considered as an extremely thin solar sail (Fig. \ref{fig:solarsail_rotation}). 
To simplify the problem, we made two main assumptions in this context. First, the radiation pressure mainly exerts 
on the maximum face, whose normal vector is along the $z$ axis, and the forces felt by other faces are ignored. 
Second, we assume the extremely thin `Oumuamua is a near-symmetric top that has a moment of inertia 
$I = (I_1, I_2, I_3) $ with $I_1 = I_2 < I_3$. 
Under the first assumption, the radiation pressure on the a perfectly reflecting light sail is
\begin{equation}
    \vec F_{\rm radiation} = { A L_{\odot} \over 2 \pi r^2 c} |{\vec e_r\cdot \vec e_z}|({\vec e_r\cdot \vec e_z}) \vec e_z.
\label{eq:fradpre}
\end{equation}
Here $A |{\vec e_r\cdot \vec e_z}| $ is the face area that is normal to the solar light and $L_{\odot}$ is the solar luminosity. 
The unit vector $\vec e_z$ is
\begin{equation}
    \vec e_z = R(\phi,\theta,\psi)  \vec e_z^\prime = 
\begin{pmatrix}
     \sin \phi \sin \theta\\\
    -\cos \phi \sin \theta\\\
     \cos \theta.
\end{pmatrix}
.\end{equation}
Without the loss of generality, we set the initial $\phi_0 = \pi/2$ so that the initial $\vec e_z$ lies on the plane ($X,\,Z$)
and $\vec e_z$ becomes
\begin{equation}
    \vec e_z = 
\begin{pmatrix}
     \sin (\phi_0+\phi) \sin \theta\\\
    -\cos(\phi_0+\phi) \sin \theta\\\
     \cos \theta
\end{pmatrix} 
= 
\begin{pmatrix}
     \cos \phi \sin \theta\\\
     \sin \phi \sin \theta\\\
     \cos \theta
\end{pmatrix} 
,\end{equation}
where $\theta$ is the nutation angle.
The unit vector of the position in the nonrotating reference frame ($XYZ$) can be assigned as $\vec e_r = (\sin \theta^\prime,\, 0,\, \cos \theta^\prime ),$ where $\theta^\prime$ is in the range ($0,\,\pi$).
The solution for a symmetric top is 
\begin{equation}
    \dot \theta = 0, \ \ \ \ \ \ {\rm and} \ \ \ \ \ \ 
    \dot \phi = {J \over I_2}.
\end{equation}

\subsection{Averaging over a spin period}
With a constant $\theta$, $\vec F_{\rm radiation}$ is a periodic force that is modulated on the timescale of the precession period $2 \pi /\dot \phi$. 
The mean force, integrating over its period, is
\begin{equation}
    \vec F_{\rm mean} = { 1 \over 2 \pi}\int_0^{2 \pi} \vec F_{\rm radiation} d\phi = { 1 \over 2 \pi}\int_0^{2 \pi} { A L_{\odot} \over 2 \pi r^2 c} |{\vec e_r\cdot \vec e_z}|({\vec e_r\cdot \vec e_z}) \vec e_z d\phi
,\end{equation}
where $\vec e_r \cdot \vec e_z 
= k_1 \cos \phi + k_2 = k_1 ( \cos \phi + k_2/k_1)$, where $k_1 = \sin \theta \sin \theta^\prime $ and $k_2 = \cos \theta \cos 
\theta^\prime $.

In the limit $k_2/k_1 \geq 1$, equivalently $\theta^\prime \leq {\pi/2} - \theta$, we have $\vec e_r \cdot \vec e_z \geq 0,$
and the mean force can be expressed as 
\begin{align}
\vec F_{\rm mean} & = { 1 \over 2 \pi}\int_0^{2 \pi} { S L_{\odot} \over 2 \pi r^2 c} ({\vec e_r\cdot \vec e_z})^2 \vec e_z d\phi \\\
& = { A L_{\odot} \over 4 \pi^2 r^2 c} \int_0^{2 \pi}(k_1 \cos \phi + k_2)^2
\begin{pmatrix}
     \cos \phi \sin \theta\\\
     \sin \phi \sin \theta\\\
     \cos \theta
\end{pmatrix} d\phi.
\end{align}
With a constant $\theta$,  
\begin{equation}
    \vec F_{\rm mean} = { A L_{\odot} \over 4 \pi r^2 c}
\begin{pmatrix}
     2k_1 k_2 \sin \theta \\\
     0 \\\
     k_1^2 + 2k_2^2.
\end{pmatrix} 
\label{eq:fmeantheta}
.\end{equation}
In the limit $k_2/k_1 \leq -1$, or equivalently $\theta^\prime \geq {\pi/2} + \theta$,  $\vec e_r \cdot \vec e_z \leq 0,$ and 
the mean force reduces to 
\begin{equation}
\begin{aligned}
    \vec F_{\rm mean} &= -{ 1 \over 2 \pi}\int_0^{2 \pi} { A L_{\odot} \over 2 \pi r^2 c} ({\vec e_r\cdot \vec e_z})^2 
    \vec e_z d\phi \\
    &=- { A L_{\odot} \over 4 \pi^2 r^2 c}
\begin{pmatrix}
     2 \pi k_1 k_2 \sin \theta \\\
     0 \\\
     \pi(k_1^2 + 2k_2^2) \cos \theta.
\end{pmatrix} 
\end{aligned}
.\end{equation}
In the intermediate region with $ -1 < k_2/k_1 < 1$,  or equivalently $  {\pi/2} - \theta<\theta^\prime 
< {\pi/2} + \theta$, we find that the condition for $\vec e_r \cdot \vec e_z > 0$ is $ -\beta <\phi < \beta,$ 
where $\beta$ is defined as 
$    \beta =  \arccos (-{k_2 \over k_1}).$
The mean force becomes
\begin{equation}
\begin{aligned}
\vec F_{\rm mean} 
& = { 1 \over 2 \pi}\int_{-\beta}^{\beta} { S L_{\odot} \over 2 \pi r^2 c} ({\vec e_r\cdot \vec e_z})^2 \vec e_z d\phi -  { 1 \over 2 \pi}\int_{- \pi}^{-\beta} { S L_{\odot} \over 2 \pi r^2 c} ({\vec e_r\cdot \vec e_z})^2 \vec e_z d\phi - 
\\ &  { 1 \over 2 \pi} \int_{\beta}^{\pi} { S L_{\odot} \over 2 \pi r^2 c} ({\vec e_r\cdot \vec e_z})^2 \vec e_z d\phi\\
& = { S L_{\odot} \over 2 \pi^2 r^2 c} \int_{0}^{\beta} (k_1 \cos \phi + k_2)^2
\begin{pmatrix}
     \cos \phi \sin \theta\\\
     0\\\
     \cos \theta
\end{pmatrix} d\phi  
 \\ &-  { S L_{\odot} \over 2 \pi^2 r^2 c}  \int_{\beta}^{\pi} (k_1 \cos \phi + k_2)^2
\begin{pmatrix}
     \cos \phi \sin \theta\\\
     0\\\
     \cos \theta
\end{pmatrix} d\phi    
\\\
 & = { A L_{\odot} \over 4 \pi^2 r^2 c}
\begin{pmatrix}
F_1(\theta,\theta') \\\
0 \\\
F_3(\theta,\theta').
\end{pmatrix}
\label{eq:fmeanbeta}
\end{aligned}
,\end{equation}
where
\begin{equation}
\begin{aligned}
F_1(\theta,\theta') &= 2k_1^2 \sin \theta (2\sin \beta - {2 \over 3} \sin^3 \beta) \\
&+ 4k_2^2 \sin \theta \sin \beta + 4 k_1k_2 \sin \theta (\beta + \sin \beta \cos \beta -{\pi \over 2}) \,,\\
F_3(\theta,\theta') &=2k_1^2 \cos \theta (\beta + \sin \beta \cos \beta -{\pi \over 2}) \\
&+ k_2^2 \cos \theta (4 \beta -2\pi) + 8k_1 k_2 \cos \theta \sin \beta)   \,.
\end{aligned}
\end{equation}
\end{document}